\documentclass{aa}  

\usepackage{graphicx}
\usepackage{txfonts}
\usepackage[dvipsnames]{xcolor}
\usepackage[figuresright]{rotating}
\usepackage{hyperref}
\hypersetup{
    colorlinks=true,
    allcolors=blue
    }
\usepackage{booktabs}
\defcitealias{2022A&A...668A.118C}{Paper I}
\newcommand{\paperIt}{\citetalias{2022A&A...668A.118C}}

\begin{document}

\title{Innovative and automated method for vortex identification}
\subtitle{II. Application to numerical simulations of the solar atmosphere}

\author{
      J.~R.~Canivete Cuissa\inst{1,2}
      \and
      O.~Steiner\inst{2,3}
}

\institute{
        Center for Theoretical Astrophysics and Cosmology, Institute for Computational Science (ICS),\\University of Zurich, Winterthurerstrasse 190, 8057 Z{\"u}rich, Switzerland
        \email{jcanivete@ics.uzh.ch}
    \and        
        Istituto ricerche solari Aldo e Cele Daccò (IRSOL), Faculty of Informatics, Università della Svizzera italiana (USI),\\ Via Patocchi 57 -- Prato Pernice, 6605 Locarno, Switzerland 
     \and 
        Leibniz-Institut f\"ur Sonnenphysik (KIS),\\ 
        Sch\"oneckstrasse 6, 79104 Freiburg i.Br., Germany
}

\date{Received 22 February 2023 / Accepted 13 November 2023}

\abstract
{Ubiquitous small-scale vortical motions are seen to occur in the solar atmosphere both in simulations and observations. They are thought to play a significant role in the local heating of the quiet chromosphere and corona. In a previous paper, we proposed a new method for the automated identification of vortices based on the accurate estimation of curvature centers; this method was implemented in the SWIRL algorithm.}
{We aim to assess the applicability of the SWIRL algorithm to self-consistent numerical simulations of the solar atmosphere. The highly turbulent and dynamical solar flow poses a challenge to any vortex-detection method.
We also conduct a statistical analysis of the properties and characteristics of photospheric and chromospheric small-scale swirling motions in numerical simulations.}
{We applied the SWIRL algorithm to realistic, three-dimensional, radiative, magneto-hydrodynamical simulations of the solar atmosphere carried out with the CO5BOLD code. In order to achieve statistical validity, we analyzed $30$ time instances of the simulation covering $2\,{\rm h}$ of physical time.}
{The SWIRL algorithm accurately identified most of the photospheric and chromospheric swirls, which are perceived as spiraling instantaneous streamlines of the horizontal component of the flow. Part of the identified swirls form three-dimensional coherent structures that are generally rooted in magnetically dominated intergranular lanes and extend vertically into the chromospheric layers. From a statistical analysis, we find that the average number densities of swirls in the photosphere and chromosphere are $1\,{\rm Mm}^{-2}$ and $4\,{\rm Mm}^{-2}$, respectively, while the average radius is $50-60\,{\rm km}$ throughout the simulated atmosphere. We also find an approximately linear correlation between the rotational speed of chromospheric swirls and the local Alfvén speed. We find evidence that more than $80\,\%$ of the identified, coherent, vortical structures may be Alfvénic in nature.}
{The SWIRL algorithm is a reliable tool for the identification of vortical motions in magnetized, turbulent, and complex astrophysical flows. It can serve to expand our understanding of the nature and properties of swirls in the solar atmosphere. A statistical analysis shows that swirling structures may be smaller, more numerous, and may rotate faster than previously thought, and also suggests a tight relation between swirls and the propagation of Alfvénic waves in the solar atmosphere.}


\titlerunning{Innovative and automated vortex identification method II}

\maketitle
   
%
%
\section{Introduction}
\label{sec:introduction}

Observations carried out over the past two decades indicate that small-scale vortical motions are ubiquitous in the quiet solar atmosphere. Many of the vortex detections have been obtained by individually following the trajectories of bright points (BPs) and small-scale magnetic structures \citep[][]{2008ApJ...687L.131B, 2010A&A...513L...6B, 2011A&A...531L...9M} 
or by visual tracking of swirling photospheric and chromospheric features, such as rings, filaments, and arcs \citep[][]{2009A&A...507L...9W, 2012Natur.486..505W, 2016A&A...586A..25P, 2018A&A...618A..51T, 2019A&A...623A.160T, 2019ApJ...881...83S}. 
Another approach is to use local correlation tracking (LCT) techniques to identify spiraling motions in the morphology of the estimated horizontal velocity fields 
\citep[][]{2010ApJ...723L.139B, 2011MNRAS.416..148V, 2017ApJS..229...14R, 2018A&A...610A..84R}. A review on vortical motions in the solar atmosphere is presented in \citet[][]{2022...ISSI}. 

These methods yielded precious results on the characteristic sizes, lifetimes, and rotational periods of photospheric and chromospheric swirls. However, they are possibly biased by the human conception of the definition of a vortex, because the detection processes rely on the visual identification of swirling motions from images, time sequences, and velocity field maps derived from observations. A more pragmatic approach consists in using mathematical criteria and geometrical methods to limit the effects of human subjectivity in the identification process. A number of such vortex identification methods can be found in the literature \citep[see, e.g.,][for a review]{2018CGF...37...6G}. 

Many of these methods have been employed to study small-scale vortical motions in the solar atmosphere, especially in the context of numerical simulations where the necessary physical quantities are directly accessible. 
For example, studies using the $\Gamma$ functions \citep{2001MeScT..12.1422G} to study photospheric vortices in observations yielded average diameters of ${\sim 0.5\,{\rm Mm}}$ and lifetimes of $\sim 0.3\,{\rm min}$ \citep[][]{2018ApJ...869..169G, 2019NatCo..10.3504L}. In the chromosphere, \citet{2022A&A...663A..94D} reported swirls with a mean diameter of $2.6\pm0.6\,{\rm Mm}$ and a lifetime of about $10.3\,{\rm min}$ using a new morphological approach \citep{2021SoPh..296...17D}. From simulations, chromospheric swirls with typical diameters of $0.7\pm0.3\,{\rm Mm}$ and average lifetimes of $\sim 1.0\,{\rm min}$ have been found by \citet[][]{2017A&A...601A.135K} using the swirling strength criterion \citep{1999JFM...387..353Z}. 
We further refer the reader to \citet[][]{1998ApJ...499..914S}, \citet[][]{2011A&A...533A.126M}, \citet[][]{2011A&A...526A...5S}, \citet[][]{2012ApJ...751L..21K}, \citet[][]{2012A&A...541A..68M}, \citet[][]{2012ASPC..456....3S}, \citet[][]{2013ApJ...776L...4S}, \citet[][]{2018ApJ...863L...2S}, \citet[][]{2019ApJ...872...22L}, \citet[][]{2020A&A...639A.118C}, \citet[][]{2020ApJ...894L..17Y}, \citet[][]{2020ApJ...898..137S}, \citet[][]{2021ApJ...915...24S}, \citet[][]{2021A&A...649A.121B}, and \citet[][]{2022ApJ...928....3A} for a nonexhaustive list of studies employing mathematical and geometrical methods to analyze swirling motions in observations and numerical simulations of the solar atmosphere. 

However, a universally accepted and rigorous method for vortex identification has not yet been found. Indeed, all the proposed methods present shortcomings when applied to the magnetized, turbulent, and highly dynamical flows of the solar atmosphere \citep[see, e.g.,][for a discussion]{2022A&A...668A.118C}. In \citet[][]{2022A&A...668A.118C} (hereafter: \citetalias{2022A&A...668A.118C}), we presented a new method for the automated identification of vortices, called the estimated vortex center (EVC) method. It combines the accuracy and quantitative aspects of mathematical criteria with the global and morphological perspective of the curvature center method proposed by 
\citet[][]{Sadar99}. 
We implemented the method in a Python package called SWirl Identification by Rotation-centers Localization (SWIRL) \citep{SWIRL}, which is open source on GitHub\footnote{\url{https://github.com/jcanivete/swirl}}. 

The SWIRL algorithm was tested on an artificial velocity field composed of nine Lamb-Oseen vortex models with a random Gaussian noise and on the turbulent flow resulting from a magneto-hydrodynamical (MHD) Orszag-Tang vortex system. In particular, the MHD Orszag-Tang test yields a flow with a diverse spectrum of MHD modes, shocks, and turbulence \citep[see, e.g.,][]{2000ApJ...530..508L}. Consequently, accurate vortex identification in such a complex flow poses a significant challenge to any dedicated algorithm. The results showed the reliability and robustness of the algorithm in the presence of noise, turbulence, and magnetic fields. Moreover, as the EVC method does not require the use of a threshold, vortices with rotational velocities that are comparable to the noisy background velocity field are not precluded from being identified. Therefore, the SWIRL algorithm proved to be suitable for identifying vortices in astrophysical velocity fields.

There are many open questions regarding small-scale swirls in the solar atmosphere. For example, the typical size, number density, strength, and lifetime of these events have been the subject of multiple observational and numerical studies. However, the obtained results most certainly depend on the spatial resolution of the instrumentation or simulation \citep[][]{2020ApJ...894L..17Y}, and on the methods employed \citep[see, e.g.,][]{2018ApJ...863L...2S, 2018SoPh..293...57T}. It is also not  yet clear whether coherent vortical structures observed in the lower solar atmosphere can extend into the corona \citep[][]{2022A&A...658A..45B}. 

Moreover, being tightly coupled to the small-scale magnetic field of the Sun, small-scale vortical motions could be associated with torsional Alfvén waves \citep[][]{2012Natur.486..505W, 2013ApJ...776L...4S}. 
Signatures of torsional Alfvén waves in the solar atmosphere have been found by, for example, \citet[][]{2009Sci...323.1582J}, \citet[][]{2011ApJ...736L..24O}, \citet[][]{2012ApJ...752L..12D}, and \citet[][]{2017NatSR...743147S}, 
while \citet[][]{2019NatCo..10.3504L} and \citet[][]{2021A&A...649A.121B} reported upwardly propagating torsional Alfvénic pulses related to chromospheric swirls in observations and numerical simulations, respectively. These studies, among others, indicate that the energy flux associated with vortical events can sustain the radiative losses in the chromosphere, and therefore these events can contribute to local heating.

The role that small-scale swirls may play in the dynamics of the solar atmosphere calls for a rigorous method for their identification. In particular, a robust statistical analysis of their properties is required to asses their real impact on chromospheric and coronal heating. In this paper, we demonstrate that the method presented in \citetalias{2022A&A...668A.118C} also reliably identifies swirls in the turbulent and highly dynamical flow of a three-dimensional, MHD numerical simulation of the solar atmosphere. Moreover, we carry out a statistical analysis on the properties and characteristics of small-scale swirling motions in those simulations.

The paper is organized as follows. In Sect.\,\ref{sec:method}, we briefly describe the numerical simulations and the vortex identification method used in this work. In Sect.\,\ref{sec:results_and_discussion}, we present and discuss the performance of the method when applied to numerical simulations of the solar atmosphere and a statistical analysis of the identified swirls. Finally, we summarize our findings and present our conclusions in Sect.\,\ref{sec:conclusions}. 

%
%

\section{Methods}
\label{sec:method}

%
%

\subsection{Numerical simulations} 
\label{subsec:numerical simulations}

\begin{figure}
        \centering
        \resizebox{\hsize}{!}{\includegraphics{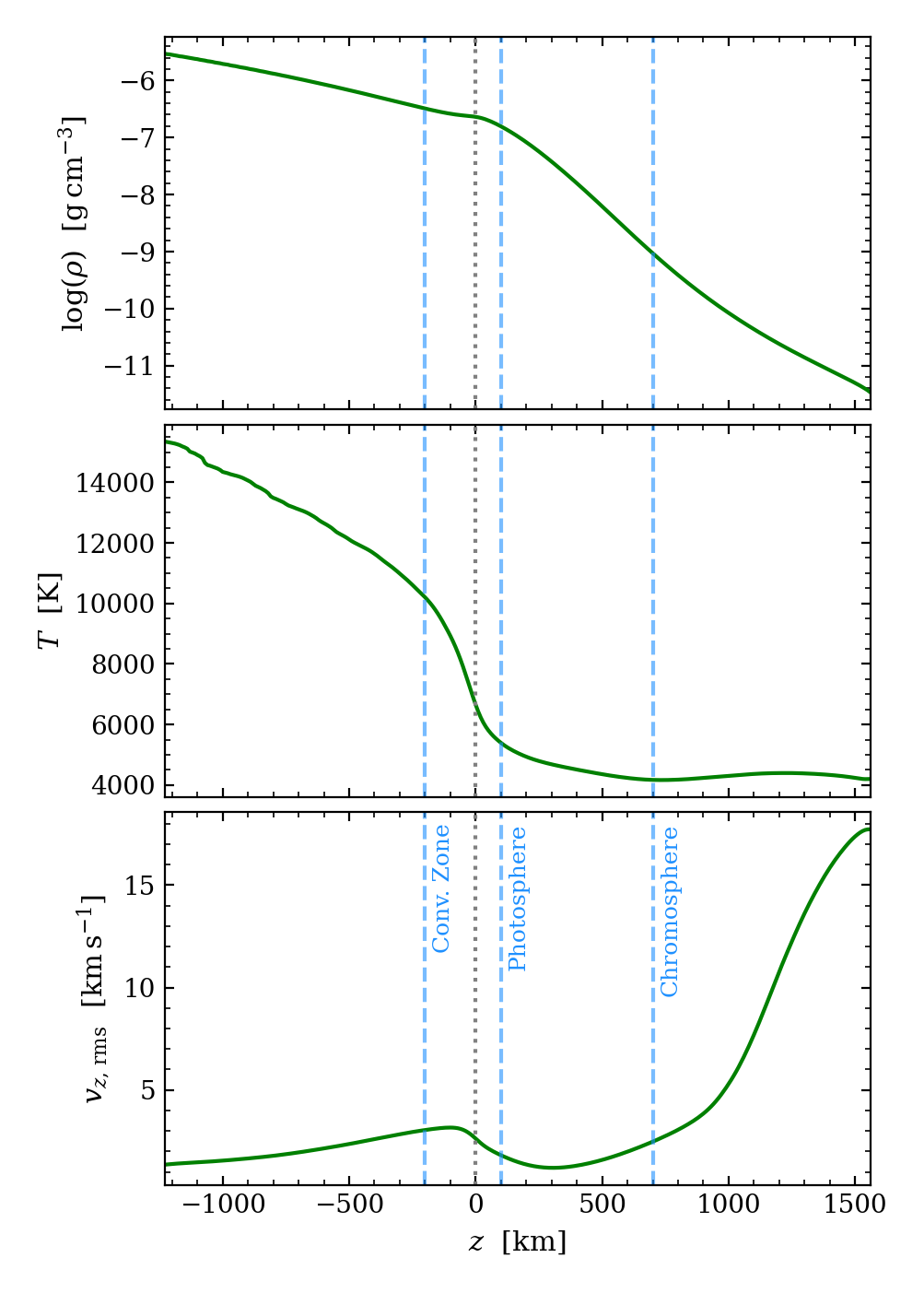}}
        \caption{Average stratifications of density, $\rho$, temperature, $T$, and rms of the vertical component of the velocity field, $v_{z,\,{\rm rms}}$. The profiles represent averages, both temporally across the 30 time instances of the CO5BOLD simulation, and spatially across the horizontal sections of the domain. The heights for the analysis of swirls in the surface layers of the convection zone ($z = -200\,{\rm km}$), in the photosphere ($z=100\,{\rm km}$), and in the low chromosphere ($z = 700\,{\rm km}$) are indicated by blue dashed lines. }
        \label{fig:sim_properties}
\end{figure}

We employed realistic numerical simulations of the solar atmosphere obtained with the radiative MHD code CO5BOLD \citep{2012JCoPh.231..919F}. 
The size of the Cartesian simulation box is $9.6\,\times9.6\,\times 2.8\,{\rm Mm}^3$ and the cell size is $10\,{\rm km}$ in each spatial direction. The number of grid cells is therefore $960\,\times960\,\times 280$. The average optical surface $\tau_{500} = 1$, which we label as $z = 0\,{\rm km}$, is found at $\sim 1300\,{\rm km}$ from the bottom of the box. Therefore, the simulation domain represents a small volume near the solar surface, which includes the surface layers of the convection zone, the photosphere, and up to the middle chromosphere. The average stratifications of density, temperature, and the root mean square (rms) of the vertical component of the velocity field are shown in Fig.\,\ref{fig:sim_properties}.

The simulation started from a relaxed, purely hydrodynamical model, to which a unipolar, vertical magnetic field of $50\,{\rm G}$ was added. The lateral boundary conditions are periodic for both the plasma and the magnetic field, while at the top and bottom of the box the magnetic field is forced to be vertically oriented. More details on the simulation setup can be found in \citet[][Sect. 2]{2018unige:115257C} 
and in \citet[][]{2021A&A...649A.121B}. 

This choice of initial magnetic field configuration is roughly representative of a predominantly unipolar magnetic network patch of a quiet Sun region. The configuration and top boundary condition of the magnetic field favor the production of vertically oriented vortex tubes in the chromosphere, as was demonstrated by \citet[][Appendix A]{2021A&A...649A.121B}. A stronger initial field would yield a more homogeneous structure of vortices, while vanishing magnetic fields would lead to less and rather isotropically distributed vortices.

For this study, we analyzed 30 time instances of the CO5BOLD simulation with a cadence of $4\,{\rm min}$, which cover a total of $2$ h in physical time. This cadence period is two-thirds of the mean granular lifetime \citep{1999ApJ...515..441H}.

%

%
%

\subsection{Identification algorithm} 
\label{subsec:identification_algorithm}

In this paper, we employed the EVC method presented in \citetalias{2022A&A...668A.118C}. It can be considered an extension of the curvature center method proposed by \citet{Sadar99}, 
where the velocity field and its derivatives are used instead of streamlines. In more detail, the method consists in accurately estimating the center of rotation of every rotating fluid particle (grid cell) from the instantaneous horizontal velocity field alone. Fluid particles belonging to the same vortex share a common axis of rotation \citep[][]{1979rdte.book..309L}, 
and therefore their estimated centers of rotation, dubbed EVCs, should cluster around the true core of the vortical structure. Consequently, vortices are identified through clusters of EVCs.

To accurately compute the EVC of any given grid cell that presents some degree of curvature in the velocity field, one has to estimate the radius of curvature and the radial direction of the local flow. For this purpose, we employ the Rortex criterion, $R$, proposed by \citet[][]{2018JFM...849..312T} and \citet{2019JHyDy..31..464W} 
. The Rortex criterion is a mathematical criterion, like the vorticity, and it is defined as
\begin{equation}
    R = \boldsymbol{\omega}\cdot\boldsymbol{u}_{\rm r} - \sqrt{\left(\boldsymbol{\omega}\cdot\boldsymbol{u}_{\rm r}\right)^2 - \lambda^2} \, , \label{eq:rortex}
\end{equation}
where $\boldsymbol{\omega}$ is the vorticity vector, $\boldsymbol{u}_{\rm r}$ is the normalized, real eigenvector of the velocity gradient tensor, and $\lambda$ is the swirling strength criterion. For more details on these quantities, we refer the reader to \paperIt. 
However, whereas the vorticity and other mathematical criteria are affected by the presence of shear flows, the Rortex criterion measures the rigid-body rotational part of the flow alone. Therefore, it is the optimal quantity to extract physical information on the curvature of the flow from the velocity field and allows unprecedented accuracy in the estimation of the center of rotation \citepalias{2022A&A...668A.118C}. 

Given a map containing all the computed EVCs, clusters indicating the presence of vortices can, in principle, be identified by eye. Nevertheless, in \citetalias{2022A&A...668A.118C}  
we proposed a modified version of the clustering by fast search and find of density peaks (CFSFDP) algorithm \citep[][]{2014Sci...344.1492R} to automatize the identification process. Moreover, a cleaning procedure is proposed to remove misidentifications caused by noise or by coherently nonspiraling curvatures in the flow.

The EVC method and the associated automated algorithm are implemented in an open-source Python package called SWIRL. For more details on the method, the clustering algorithm, and the test cases, we refer the reader to \citetalias{2022A&A...668A.118C}.

%
%

\section{Results and discussion}
\label{sec:results_and_discussion}

In this section, we test the applicability of the SWIRL algorithm in automatically identifying swirls in photospheric and chromospheric horizontal sections of the simulation introduced in Sect.\,\ref{subsec:numerical simulations}. We then present the results of a statistical study performed over the full set of data cubes that  addresses the properties of small-scale swirls and their relation with the surface magnetic field of the Sun in numerical simulations.

The SWIRL algorithm requires careful tuning of several parameters based on the specific characteristics of the flows being analyzed. Detailed descriptions of these parameters can be found in \paperIt ~and in the GitHub repository for the SWIRL code. The values used in this study are listed in Table\,\ref{tab:swirl_params}.

We find the identification process to be particularly sensitive to the number of ``stencils'' used, as well as the ``noise'' and ``kink'' parameters. Increasing the number of stencils increases the robustness of the identification process to small-scale turbulence. However, using too many stencils (typically more than $\sim 10$) can lead to lower computational performance without significantly improving the results. 

The noise and kink parameters are responsible for cleaning up false detections, and their adjustment depends on the level of noise and turbulence in the flow. Higher values can lead to false detections, while excessively low values can cause true vortices to be missed. Empirically, the parameter values that have shown good performance in CO5BOLD simulations of the solar atmosphere range from about $0.5$ to $1.5$. However, we encourage users of the SWIRL algorithm to experiment with different values. 

The clustering parameters can also be adjusted based on decision graphs (see Fig.\,6 in \paperIt). However, the values given in the Table\,\ref{tab:swirl_params} should generally lead to satisfactory results for most applications.

\begin{table}[]
\renewcommand\arraystretch{1.2}
\centering
\caption{SWIRL algorithm parameters used in this work.}
\begin{tabular}{p{0.2\textwidth}p{0.08\textwidth}}
\hline\hline 
\multicolumn{2}{c}{\bfseries Criterion} \\ \midrule
Stencils                 & $1,2,3,4,5,7$       \\
$\epsilon_{\lambda}$     & $0.0$           \\
$\kappa_{\lambda}$       & $0.9$          \\
$\delta_{\lambda}$       & $0.9$           \vspace{0.2cm} \\ \hline
\multicolumn{2}{c}{\bfseries Clustering} \\ \midrule
$d_{\rm c}$              & $100\,{\rm km}$ \\
Adaptive $d_{\rm c}$     & False           \\
Fast clustering          & True            \\
Kernel                   & Gaussian        \\
Decision method          & gamma           \\
$p_{\rho}$               & 0.9             \\
$p_{\delta}$             & 0.9             \\
$p_{\gamma}$             & 1.01            \vspace{0.2cm}  \\ \hline
\multicolumn{2}{c}{\bfseries Noise} \\ \midrule
Noise parameter          & 1.3             \\
Kink parameter           & 0.5  \vspace{0.15cm}  \\ \hline
\end{tabular}
\tablefoot{More details on the role of the different parameters can be found in \citetalias{2022A&A...668A.118C} and on the GitHub repository of the code.}
\label{tab:swirl_params}
\end{table}

%
%

\subsection{Validation of the SWIRL algorithm on CO5BOLD simulations}
\label{subsec:validation}

\subsubsection{Photosphere}
\label{subsubsec:photosphere}

\begin{figure}[!ht]
        \centering
        \resizebox{\hsize}{!}{\includegraphics{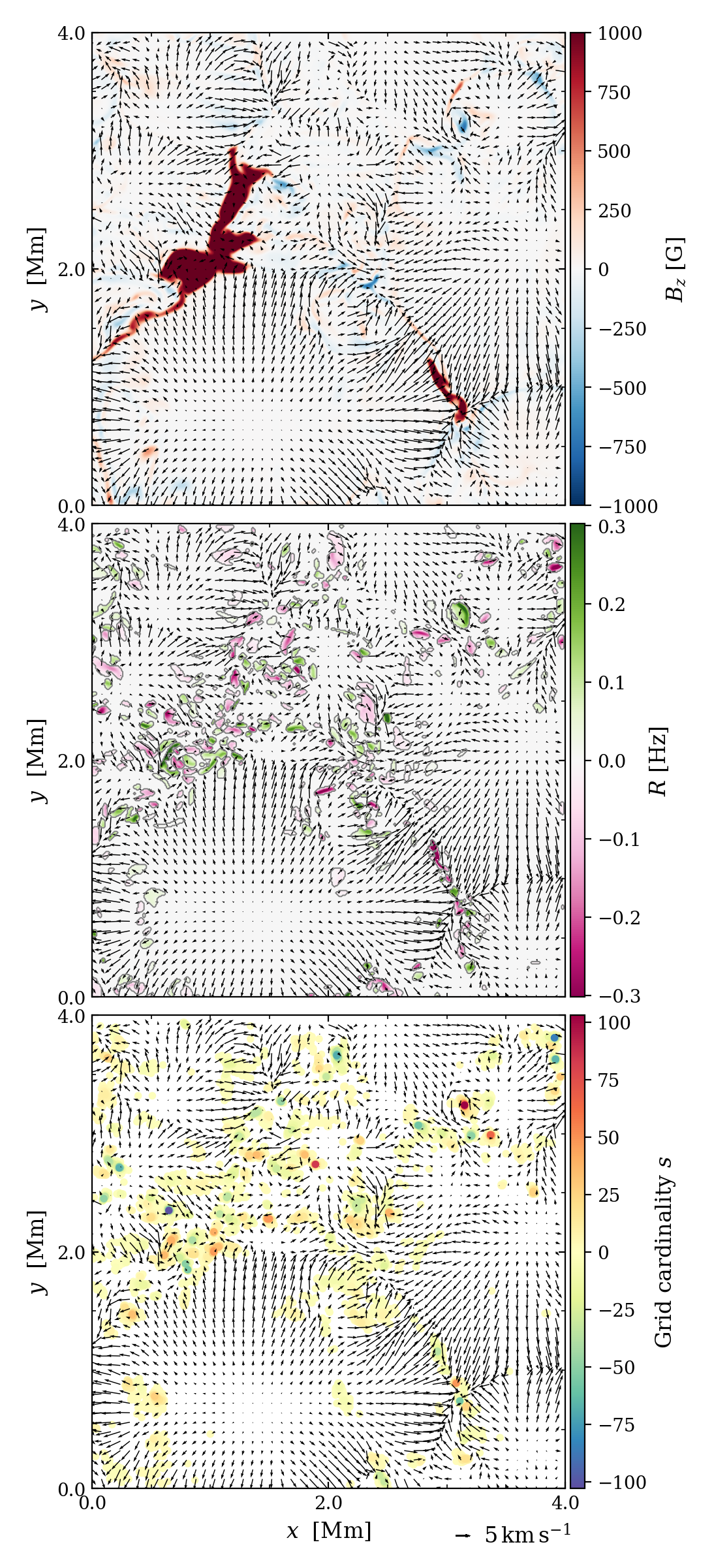}}
        \caption{Two-dimensional horizontal subsection of the simulated photosphere. The section measures $4.0\,\times\,4.0\,{\rm Mm}^2$ and is taken at $z=100\,{\rm km}$. The simulation time instance corresponds to $t=5774\,{\rm s}$. The horizontal velocity field of the subsection is depicted using a vector plot. The length of each arrow corresponds to the magnitude of the horizontal flow, and a reference scale is included in the bottom-right corner. Top: Vertical magnetic field $B_z$ at $z=100\,{\rm km}$. Middle: Rortex criterion $R$. Bottom: G-EVC map.  Contours where $R \neq 0$ are shown in gray in the middle panel.}
        \label{fig:test_sim_photosphere_map}
\end{figure}
\begin{figure}
        \centering
        \resizebox{\hsize}{!}{\includegraphics{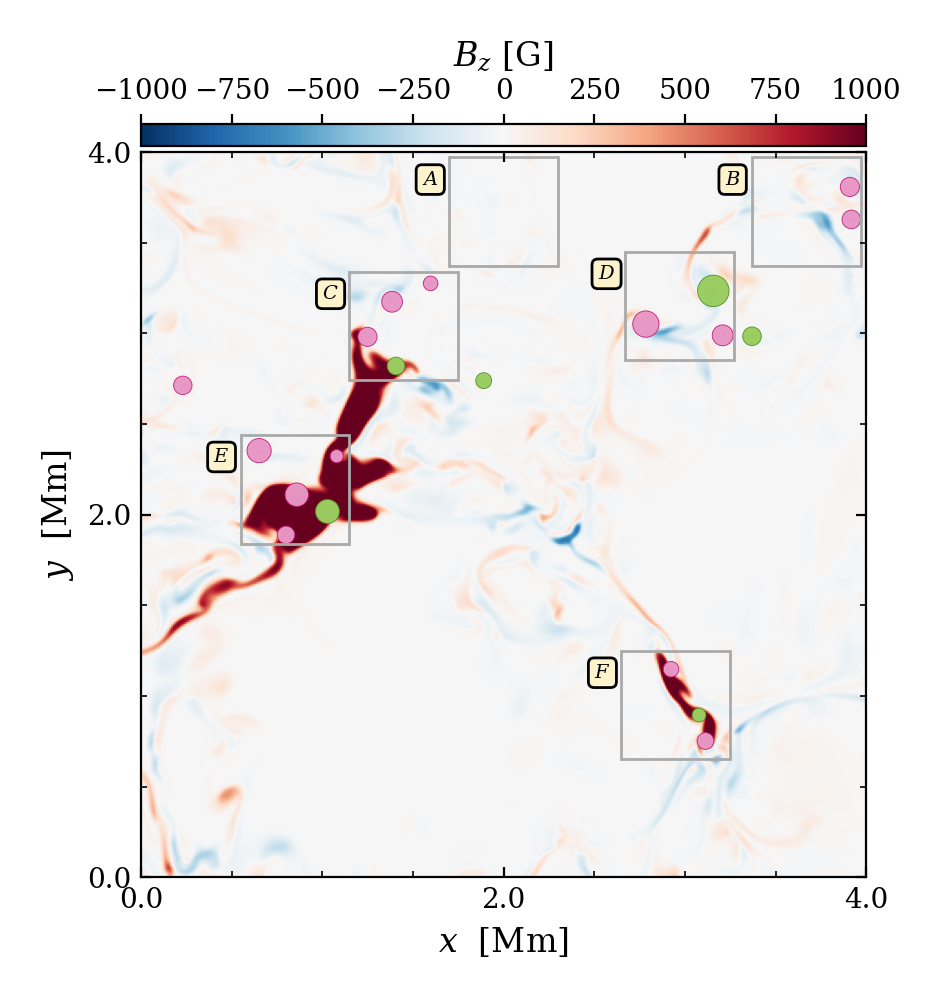}}
        \caption{Vortices identified by the SWIRL algorithm in the two-dimensional, horizontal velocity field of Fig.\,\ref{fig:test_sim_photosphere_map}. The location and effective size of the identified vortices are indicated by colored disks. Clockwise vortices are  represented by purple disks, while counterclockwise ones are shown in green. The vertical magnetic field $B_z$ is color coded and saturates at $\pm 1000\,{\rm G}$. The gray squares denote the $0.6\times 0.6\,{\rm Mm}^2$ regions shown in Fig.\,\ref{fig:test_sim_photosphere_zoomin}.}
        \label{fig:test_sim_photosphere}
\end{figure}
\begin{figure}
        \centering
        \resizebox{\hsize}{!}{\includegraphics{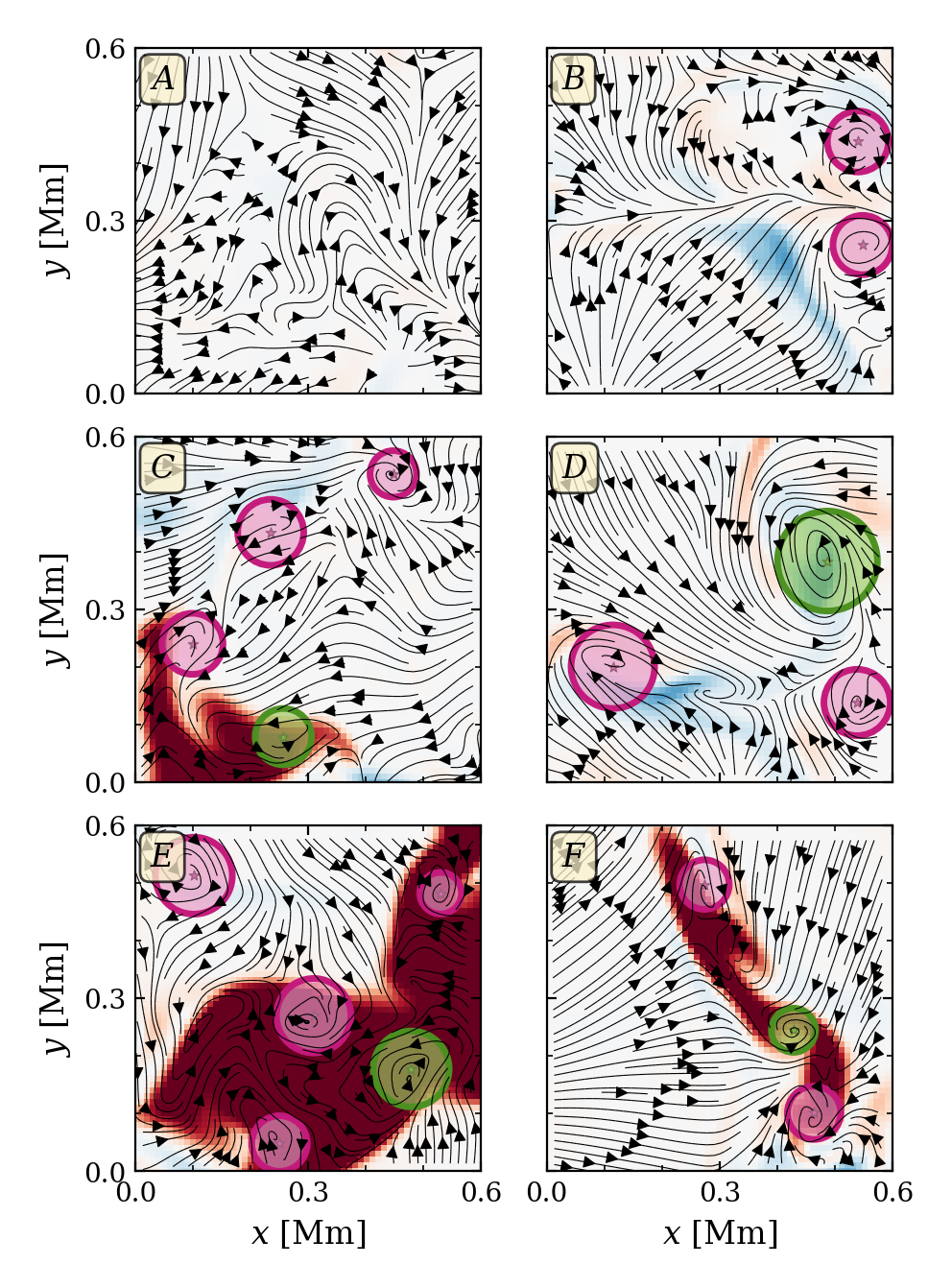}}
        \caption{Zoom-in plots of the photospheric regions outlined in Fig.\,\ref{fig:test_sim_photosphere}. The location and effective size of the identified vortices are indicated by colored disks. The vertical magnetic field $B_z$ is color coded with the same scale as in Fig.\,\ref{fig:test_sim_photosphere}, while the horizontal velocity field is represented by instantaneous streamlines.}
        \label{fig:test_sim_photosphere_zoomin}
\end{figure}

We started from a photospheric, two-dimensional, horizontal subsection of a time instance of the simulation data, which is shown in Fig.\,\ref{fig:test_sim_photosphere_map}. The chosen height is $z=100\,{\rm km}$ as shown in Fig.\,\ref{fig:sim_properties}. We notice the granular pattern of the flow with integranular lanes harboring magnetic flux concentrations (top panel). The magnetic field is predominantly of positive polarity because of the initial condition of the simulation. 

The middle panel of Fig.\,\ref{fig:test_sim_photosphere_map} shows the Rortex criterion, $R$, computed from the horizontal velocity field. Positive values of $R$ (green) indicate counterclockwise curvature in the flow, while clockwise curvatures are characterized by negative $R$ (purple). Small-scale patches where $R\neq0$ appear to roughly track the photospheric magnetic flux concentrations, but form a rather chaotic pattern of different rotational strengths and orientations\footnote{The strength of the Rortex criterion is proportional to the rigid-body angular velocity of the flow, while its sign describes the orientation according to the right-hand rule.}. 

It would be difficult (if not impossible) to discern coherent vortical structures using the Rortex map alone. A priori, we do not know if a two-dimensional region of $R\neq0$ is part of a coherent vortical structure or simply stems from the turbulent nature of the flow. Indeed, the Rortex criterion is a local criterion, defined on a small stencil of only very few grid cells. However, to distinguish turbulent, local rotations from actual vortical flows, additional information about the large-scale properties of the flow is needed. 
For example, if a single fluid parcel is deflected, it may exhibit local rotation and therefore be identified by a mathematical criterion such as Rortex. However, the presence of a vortex flow requires several fluid parcels to rotate coherently about a common axis.  It is the coherent behavior of many parcels that indicates the presence of a vortical structure.

This conundrum can be partially solved by considering the G-EVC map, which is shown in the bottom panel of Fig.\,\ref{fig:test_sim_photosphere_map}. The G-EVC map is obtained by counting the number of EVCs in every grid cell.\footnote{While each grid cell yields one single EVC, the resulting coordinates of more than one EVC can fall into one and the same grid cell, giving rise to the grid cardinality $s$ of this cell.} 
Clockwise and counterclockwise EVCs count as $-1$ and $+1$, respectively, and their sum determines the grid cardinality, $s$, in each grid cell. In principle, a cluster of EVCs indicates the location of a vortex core, and therefore high absolute values of the grid cardinality, $|s|$, can be used to infer the presence of a vortex. For example, by inspecting the bottom panel of Fig.\,\ref{fig:test_sim_photosphere_map}, we expect a counterclockwise vortex to be found around $(x,\,y) = (3.1\,{\rm Mm},\,3.25\,{\rm Mm})$ and a clockwise one close to $(x,\,y) = (0.7\,{\rm Mm},\,2.4\,{\rm Mm})$. On the other hand, we can presume that the Rortex patches around $(x,\,y) = (0.2\,{\rm Mm},\,3.7\,{\rm Mm})$ visible in the middle panel of Fig.\,\ref{fig:test_sim_photosphere_map} do not represent a swirl, because the grid cardinality is relatively low in that region. 

The SWIRL algorithm automatically finds clusters of G-EVCs, and thus detects candidate vortex centers. The vortices identified on the $4.0 \times 4.0\,{\rm Mm}^2$ photospheric subsection of the CO5BOLD simulation are shown in Fig.\,\ref{fig:test_sim_photosphere}. For simplicity, we represent vortices with colored disks centered on the estimated vortex core. However, it is important to note that the SWIRL algorithm returns a collection of grid cells that form the vortex for each identification. As a result, the true shape of the identified vortices, while generally exhibiting a roundish appearance, tends to be more irregular than what is presented here. The radius of the disk corresponds to the effective radius of the vortex, $r_{\rm eff}$, which is computed as
\begin{equation}
    r_{\rm eff} = \sqrt{\frac{N_{\rm c}}{\pi}}\Delta x \, \label{eq:effective_radius}
,\end{equation}
where $N_{\rm c}$ is the number of EVCs belonging to the cluster and $\Delta x$ is the grid spacing. Here, the effective radius is defined through the effective area occupied by the grid cells that form that vortex.
The color of the disks indicates the rotation direction: green for counterclock wise vortices and purple for clockwise ones. In total, 21 vortices have been identified by the code with an average effective radius of $\sim 50\,{\rm km}$.

Most of the detected vortices lie within or nearby strong magnetic flux concentrations. This is expected, because photospheric swirling motions are known to be tightly coupled to small-scale surface magnetic fields \citep[see, e.g.,][]{2012A&A...541A..68M, 2021A&A...649A.121B}. 
There are also a few exceptions: for example, the two clockwise vortices around $(x,\,y) = (4.0\,{\rm Mm},\,3.7\,{\rm Mm})$. These apparently nonmagnetic events could be related to the footpoints of vortex arches in high-plasma-$\beta$ regions or to nonmagnetic bright points, such as those reported in numerical simulations by \citet[][]{2010NewA...15..460M}, \citet[][]{2011A&A...533A.126M}, \citet[][]{2021A&A...649A.121B}, 
and \citet[][]{2016A&A...596A..43C}.

Figure \ref{fig:test_sim_photosphere_zoomin} shows zoom-in plots of six different $0.6\times0.6\,{\rm Mm}^2$ regions of the photospheric section shown in Fig.\,\ref{fig:test_sim_photosphere}. The horizontal velocity field, which is represented by instantaneous streamlines, is particularly turbulent in magnetic flux concentrations, resulting in multiple spiraling configurations within the same magnetic structure. In general, the identified vortices correlate well with the spiraling instantaneous streamlines. 

In panel A, no vortices have been identified despite the large negative value of the grid cardinality, $s$, in that same region (see bottom panel of Fig.\,\ref{fig:test_sim_photosphere_map}). That cluster of EVCs is caused by the semi-circular clockwise configuration of the flow visible at coordinates $(x,\,y) = (0.3\,{\rm Mm},\,0.3\,{\rm Mm})$ in the center of panel A of Fig.\,\ref{fig:test_sim_photosphere_zoomin}. In this case, the SWIRL algorithm identifies the cluster of (G-)EVCs as a possible candidate vortex, but correctly discards it during the cleaning procedure because the flow is not fully spiraling. For details on the cleaning procedure, we refer the reader to \paperIt.   

The only two misidentifications are found in panel C at coordinates $(x,\,y) = (0.1\,{\rm Mm},\,0.25\,{\rm Mm})$ and $(x,\,y) = (0.25\,{\rm Mm},\,0.45\,{\rm Mm})$. The SWIRL algorithm identified two clockwise vortices in these locations. However, the instantaneous velocity streamlines do not indicate the presence of spiraling flows. Generally, the code proved to be reliable in identifying vortical motions at the photospheric level. Out of 20 identified swirls in the snapshot of Fig.\,\ref{fig:test_sim_photosphere_map}, only two were misidentified, giving an estimated accuracy of $\sim\,90\,\%$.  

%
%
\subsubsection{Chromosphere}
\label{subsubsec:chromosphere}

\begin{figure}
        \centering
        \resizebox{\hsize}{!}{\includegraphics{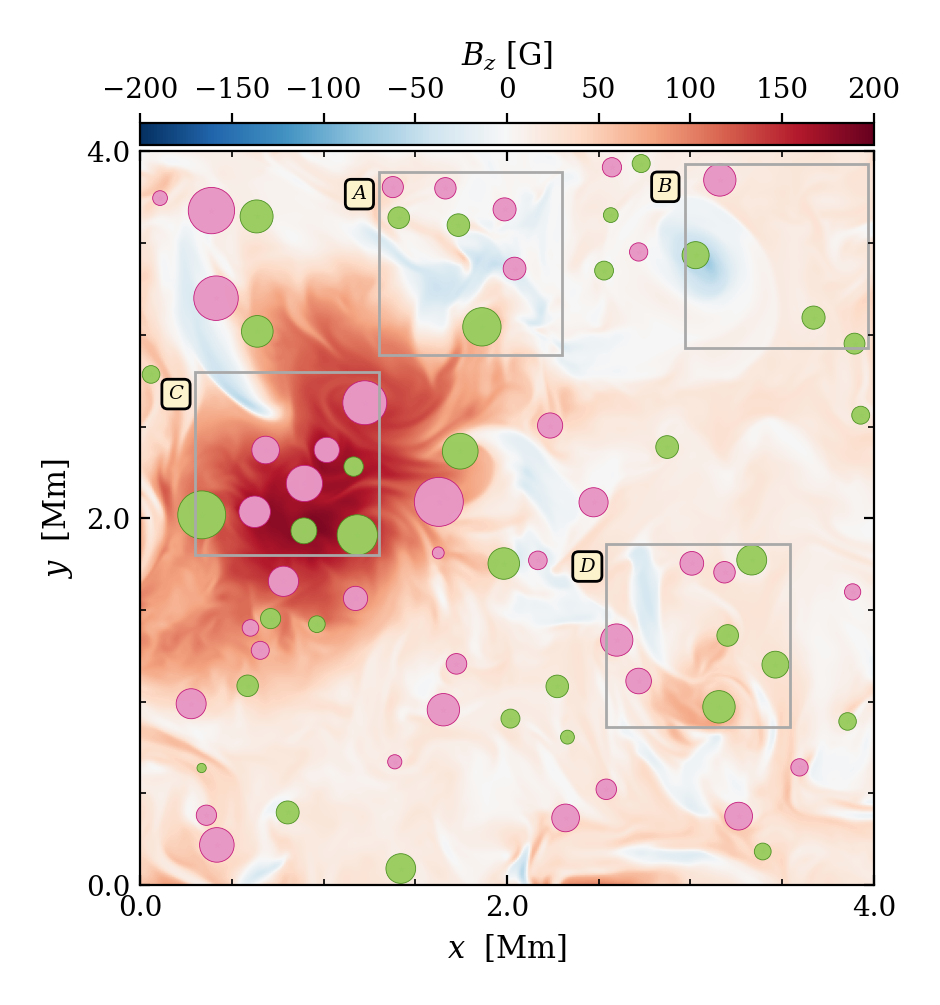}}
        \caption{Vortices identified by the SWIRL algorithm in the same horizontal domain as that of Fig.\,\ref{fig:test_sim_photosphere} but at the base of the chromosphere ($z = 700\,{\rm km}$). The location and effective size of the identified vortices are indicated by colored disks. Clockwise vortices are shown in purple, while counter-clockwise ones are shown in green. The vertical magnetic field $B_z$ is color coded and saturated at $\pm 200\,{\rm G}$. The gray squares denote the $0.6\times 0.6\,{\rm Mm}^2$ regions shown in Fig.\,\ref{fig:test_sim_chromosphere_zoomin}.}
        \label{fig:test_sim_chromosphere}
\end{figure}
\begin{figure}
        \centering
        \resizebox{\hsize}{!}{\includegraphics{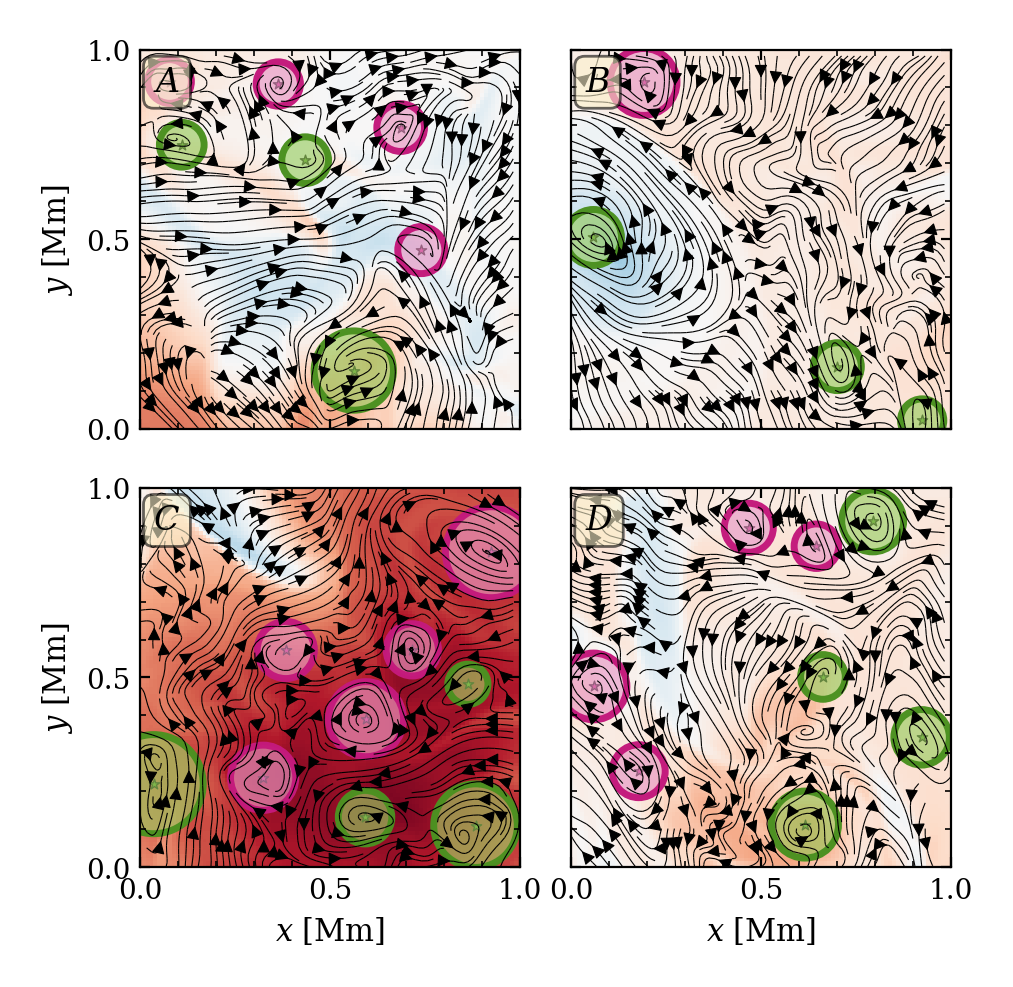}}
        \caption{ Zoom-in plots of the chromospheric regions outlined in Fig.\,\ref{fig:test_sim_chromosphere}. The location and effective size of the identified vortices are indicated by colored disks. The vertical magnetic field $B_z$ is color coded, while the horizontal velocity field is represented by instantaneous streamlines.}
        \label{fig:test_sim_chromosphere_zoomin}
\end{figure}

To further assess the reliability of the SWIRL algorithm when applied to realistic numerical simulations of the solar atmosphere, we repeated the identification analysis on a chromospheric section of the simulation box. The chosen subsection covers the same horizontal domain at the same time instance as taken in Sect.\,\ref{subsubsec:photosphere}, but at $z=700\,{\rm km}$ above the average surface of optical depth $\tau_{500} = 1$, which corresponds to the bottom of the chromosphere (see Fig.\,\ref{fig:sim_properties}). The identified vortices are shown in Fig.\,\ref{fig:test_sim_chromosphere}, and are also shown in more detail in the zoom-in plots of Fig.\,\ref{fig:test_sim_chromosphere_zoomin}.

At first sight, we notice that the swirls identified in the chromospheric layer appear to be more numerous and larger than the photospheric ones. Indeed, in Fig.\,\ref{fig:test_sim_chromosphere} there are $74$ vortices, with the largest one measuring $266\,{\rm km}$ in diameter. Multiple swirls are found in the magnetic region around $(x,\,y) = (1.0\,{\rm Mm},\,2.0\,{\rm Mm}$), which stems from the strong photospheric magnetic flux concentration visible at the same coordinates in Fig.\,\ref{fig:test_sim_photosphere}. Panel C of Fig.\,\ref{fig:test_sim_chromosphere_zoomin} shows a $1.0 \times  1.0\,{\rm Mm}^2$ close-up view of that region with streamlines derived from the horizontal velocity field and multiple spiraling patterns of different orientation can be seen. \citet{2021A&A...649A.121B} 
found that multiple swirls typically coexist in strong and complex magnetic flux concentrations in numerical simulations, dubbing this type of formation ``superposition of swirls''. 

Overall, the SWIRL algorithm identified most of the swirls in the chromospheric section of the simulation, as we can infer from the horizontal velocity field streamlines shown in Fig.\,\ref{fig:test_sim_chromosphere_zoomin}. The estimated effective radii also correlate well with the visual size of the spiraling streamlines. There are nonetheless a few exceptions. For example, a small-scale clockwise vortex at $(x,y) \sim (0.6\,{\rm Mm},\,0.35\,{\rm Mm})$ of panel D appears to have been missed, while a misidentification probably occurred around $(x,y) \sim (0.65\,{\rm Mm},\,0.85\,{\rm Mm})$ of the same panel. Moreover, the radius of the relatively large counterclockwise vortical system shown in the left of panel B is presumably underestimated. An analysis of the radial profile of the tangential velocity \citep[as done by, e.g.,][]{2020ApJ...898..137S} would be necessary to draw robust conclusions, but based on visual inspection of closed instantaneous streamlines, we can estimate that the size of the vortex, as estimated visually, could be as much as four times larger than that computed by SWIRL.

\begin{figure*}
        \centering
        \resizebox{\hsize}{!}{\includegraphics{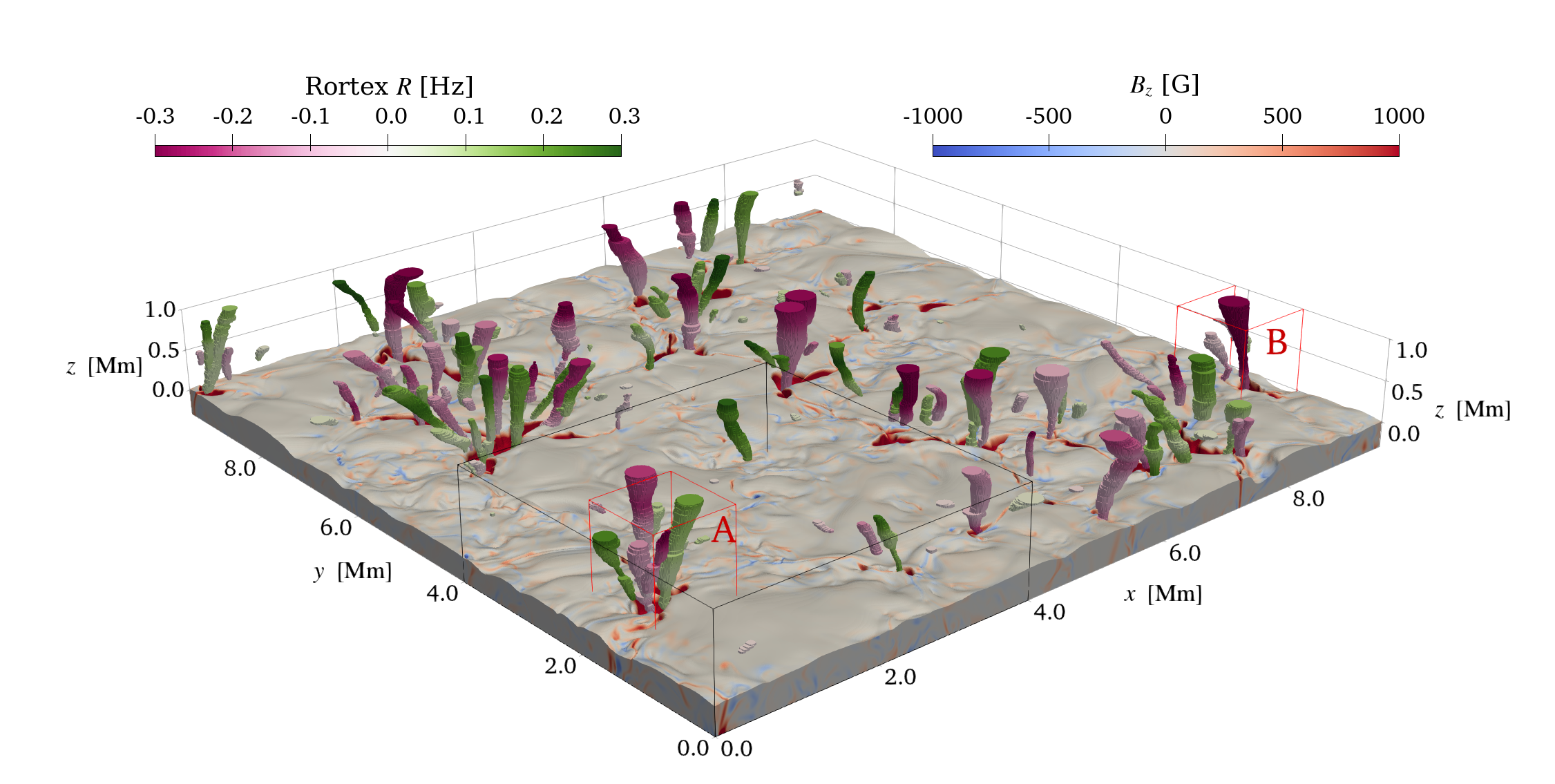}}
        \caption{Three-dimensional vortical structures identified for the time instance $t=5774\,{\rm s}$ of the CO5BOLD simulation. The displayed structures are obtained by stacking two-dimensional vortices in different height levels that are sufficiently well aligned with each other in the vertical direction. Only those structures that are rooted in the photosphere and reach the plane at $z=700\,{\rm km}$ are displayed. The three-dimensional vortices are colored according to the Rortex value $R$ averaged over their surface at each height $z$. The surface of optical depth $\tau_{500} = 1$ is shown with the vertical magnetic field $B_z$ color coded on it. The black box outlines the $4.0\times4.0\,{\rm Mm}^2$ horizontal domain used for Figs.\,\ref{fig:test_sim_photosphere} and \ref{fig:test_sim_chromosphere}, while zoomed-in renderings of the two red boxes labeled A and B are shown in Figs.\,\ref{fig:test_sim_vortexA} and \ref{fig:test_sim_vortexB}, respectively.}
        \label{fig:test_sim_3dstructures}
\end{figure*}

\subsubsection{Three-dimensional structures}
\label{subsec:3d_structures}

\begin{figure*}
        \centering
        \resizebox{\hsize}{!}{\includegraphics{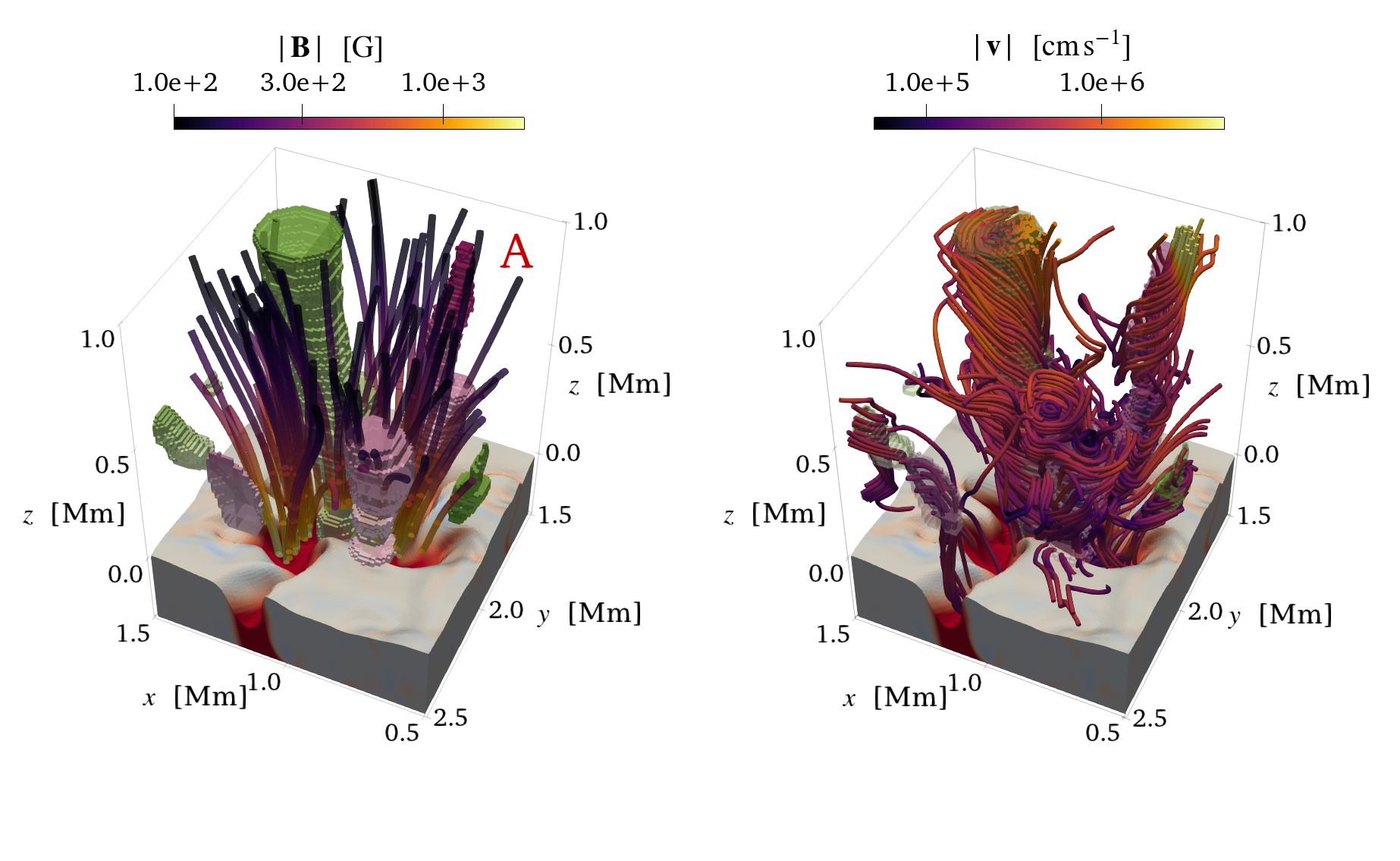}}
        \caption{Three-dimensional rendering of a superposition of swirls stemming from a relatively large and complex small-scale photospheric magnetic flux concentration. Left: Identified three-dimensional swirls colored according to the mean Rortex value $R,$ as in Fig.\,\ref{fig:test_sim_3dstructures}. Thick tubes represent magnetic field lines with the intensity of the magnetic field color coded on them. The corrugated surface near $z=0\,{\rm km}$ represents the $\tau_{500} = 1$ surface. Right: Instantaneous streamlines of the velocity field belonging to the vortical structures. The strength of the velocity field is color coded on the streamlines.}
        \label{fig:test_sim_vortexA}
\end{figure*}
\begin{figure*}
        \centering
        \resizebox{\hsize}{!}{\includegraphics{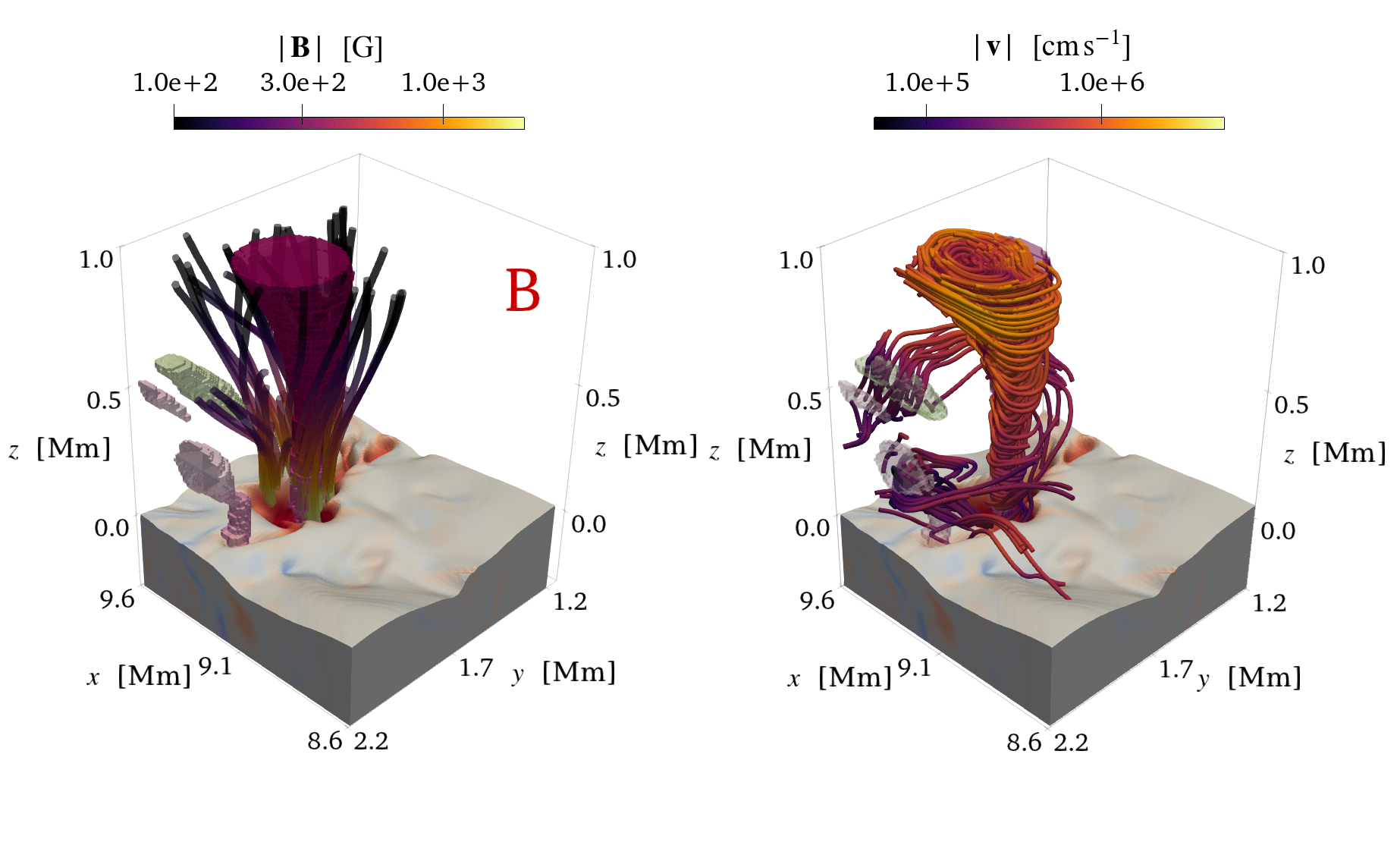}}
        \caption{Three-dimensional rendering of an isolated swirling structure stemming from a relatively small photospheric magnetic flux concentration. Left: Identified three-dimensional swirls colored according to the mean Rortex value $R,$ as in Fig.\,\ref{fig:test_sim_3dstructures}. Thick tubes represent magnetic field lines with the intensity of the magnetic field color coded on them. The corrugated surface near $z=0\,{\rm km}$ represents the $\tau_{500} = 1$ surface. Right: Instantaneous streamlines of the velocity field belonging to the vortical structures. The strength of the velocity field is color coded on the streamlines.}
        \label{fig:test_sim_vortexB}
\end{figure*}

To investigate the three-dimensionality of the vortical structures self-consistently emerging in the simulation, we applied the SWIRL algorithm to the full $9.6\times9.6\,{\rm Mm}^2$ horizontal domain at all heights between $z=-300\,{\rm km}$ (surface layers of the convection zone) and $z=1\,000\,{\rm km}$ (middle chromosphere). For this analysis, we used the same parameters (Table\,\ref{tab:swirl_params}) at all heights. 

As the automated identification is carried out on two-dimensional horizontal slices, only vertically extending vortices will be identified by our approach. Horizontal small-scale swirls have also been observed in the solar atmosphere \citep[see, e.g.,][]{2010ApJ...723L.180S, 2020ApJ...903L..10F}, but they probably do not impact the upward transport of energy and mass as they do not reach the upper atmospheric layers. 

To construct three-dimensional swirling structures, we search for vertical alignments between two-dimensional swirls identified at different heights in the simulation box. 
For this purpose, we consider two swirls with the same orientation to be part of the same vortical structure if the distance between their centers is smaller than a certain threshold. For this study, we chose the threshold to be $40\,{\rm km}$ in the horizontal direction over a vertical distance of $20\,{\rm km}$, which corresponds to four grid cells horizontally and two grid cells vertically. 
In this way, a missed identification in one plane between two adjacent planes with corresponding identification does not preclude the identification of the full three-dimensional structure. 

Moreover, as the horizontal threshold is smaller than the swirl average radius (see Sect.\,\ref{subsec:statistics}), the risk of two swirls being improperly connected is minimal. Using larger thresholds would increase the risk of erroneously connecting two separate swirls. Using excessively small thresholds carries the danger of missing a three-dimensional structure when the SWIRL algorithm misses the detection of a vortex in a single plane. 

We started with the two-dimensional swirls identified in the horizontal plane located at $z=700\,{\rm km}$. We then looked for horizontally aligned swirls in the plane $20\,{\rm km}$ below and above it. Whenever such an alignment was found, we reiterated the process starting from the previously connected two-dimensional swirl. In this way, we can construct swirls of coherent vertical extension that represent the three-dimensional extension of the two-dimensional swirls identified by the SWIRL algorithm on the different horizontal planes.

The three-dimensional vortices  identified as above are shown in Fig.\,\ref{fig:test_sim_3dstructures} for the time instance $t = 5774\,{\rm s}$ of the simulation. We note that only the vortices reaching the height of $z=700\,{\rm km}$ are displayed in this figure, because this was the starting point for the three-dimensional stacking process. Vortical structures that are restricted to the surface layers of the convection zone or photosphere  are omitted, as are purely chromospheric vortices that do not extend to the photosphere. Therefore, Fig.\,\ref{fig:test_sim_3dstructures} shows only three-dimensional swirls that connect the photosphere to the chromosphere.

The majority of the vertically extending swirls stem from photospheric magnetic flux concentrations, as we can see from the vertical magnetic field $B_z$ color coded on the $\tau_{500} = 1$ surface of Fig.\,\ref{fig:test_sim_3dstructures}. Moreover, multiple swirls coexist in strong and complex magnetic foot points, which is in agreement with the results obtained in Figs.\,\ref{fig:test_sim_photosphere} and \ref{fig:test_sim_chromosphere} from the two-dimensional sections. 

Figure \ref{fig:test_sim_vortexA} shows an example of a superposition of swirls in more detail. The three-dimensional domain, which encloses the large magnetic flux concentration located at $(x,y) \sim (1.0\,{\rm Mm},\,2.0\,{\rm Mm})$ in Figs.\,\ref{fig:test_sim_photosphere} and \ref{fig:test_sim_chromosphere}, is outlined by the red box labeled A in Fig.\,\ref{fig:test_sim_3dstructures}. Multiple vortices are identified in this patch and can be visualized by the instantaneous velocity field streamlines shown in the right panel of Fig.\,\ref{fig:test_sim_vortexA}. The magnetic field lines shown in the left panel are mostly vertically oriented. This is typical in strong magnetic flux concentrations with plasma-$\beta \ll 1$, where plasma-$\beta$ is the ratio between the gas pressure, $p_{\rm g}$, and the magnetic pressure, $p_{\rm m} = B^2/8 \pi$. We recall that swirling motions and essentially untwisted, vertically oriented magnetic fields are not mutually exclusive. Such a configuration can be thought of as a quasi-rigidly rotating, stiff magnetic flux concentration.

Also visible in Fig.\,\ref{fig:test_sim_3dstructures} are isolated swirls stemming from relatively small and weak magnetic footpoints. An example of such an event is shown in Fig.\,\ref{fig:test_sim_vortexB}, which is outlined by the red box labeled $B$ in Fig.\,\ref{fig:test_sim_3dstructures}. The magnetic field is weaker in this case; hence the plasma-$\beta$ is closer to unity in that region, in particular in the photospheric layers. Under these conditions, the flow dominates the magnetic field and the frozen-in magnetic field lines are dragged by the rotating plasma. Indeed, we observe slightly twisted magnetic field lines in the proximity of the three-dimensional swirl. The orientation of the twist in the magnetic field lines (counterclockwise) is contrary to the rotation of the flow (clockwise) when thinking of upwardly directed positive polarity. Such a configuration is compatible with the propagation of an Alfvénic pulse \citep[see ][]{2019NatCo..10.3504L, 2021A&A...649A.121B} and the event shown in Fig.\,\ref{fig:test_sim_vortexB} is structurally similar to the one analyzed by \citet[][]{2021A&A...649A.121B}, which proved to be Alfvénic in nature.

%
%
\subsection{Statistics}
\label{subsec:statistics}

In this section, we investigate the properties of small-scale swirls in the simulated solar atmosphere from a statistical point of view. For this purpose, we ran the SWIRL algorithm on the horizontal planes of $30$ time instances of the CO5BOLD simulation, covering a total physical time interval of $2$ hours. We used the full $9.6\times9.6\,{\rm Mm}^2$ horizontal extent of the simulation domain between $z = -300\,{\rm km}$ and $z = 1 000\,{\rm km}$. Our statistical analysis assesses the properties of swirls in the surface layers of the convection zone, the photosphere, and the low chromosphere of the numerical simulation. We note that all the swirls identified in the $30$ time instances have been taken into account for the analysis presented in this section, regardless of whether or not they are part of a three-dimensional structure. 

\subsubsection{Vertical profile of swirl properties}
\label{subsubsec:vertical_distribution_of_the_main_properties}

\begin{figure}
        \centering
        \resizebox{\hsize}{!}{\includegraphics{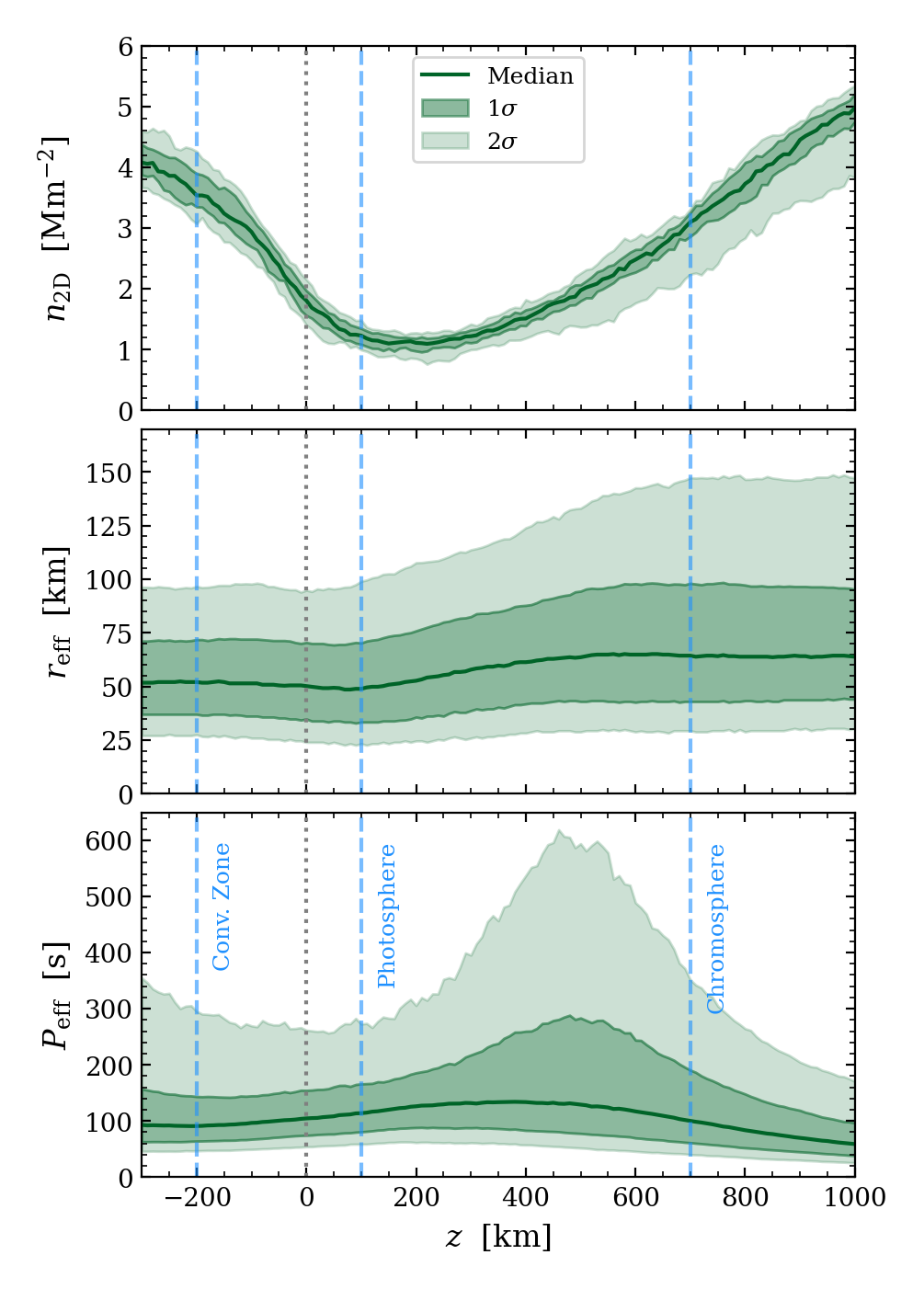}}
        \caption{ Statistical distributions as a function of height $z$ of the number density of swirls per unit area, $n_{\rm 2D}$ (top), the effective radius, $r_{\rm eff}$ (middle), and the effective period of rotation, $P_{\rm eff}$ (bottom). The median and the $1\sigma$ and $2\sigma$ deviations of the distributions are shown at each height $z$. The average optical surface $\tau_{500} = 1$ ($z=0\,{\rm km}$) is marked by a dotted line, while the heights of the surface layers of the convection zone, photosphere, and low chromosphere used in the analysis are indicated by dashed blue lines. }
        \label{fig:test_stat_z}
\end{figure}

The distributions of the number density per unit area, $n_{\rm 2D}$, the effective radius, $r_{\rm eff}$, and the effective rotational period, $P_{\rm eff}$, of the identified swirls as a function of height, $z$, are shown in Fig.\,\ref{fig:test_stat_z}. The data at each height $z$ are generally not symmetrically distributed and are best fitted by a generalized extreme value distribution. Therefore, we show the $68.2\,\%$ and $95.4\,\%$ percentile areas around the median, labeled $1\sigma$ and $2\sigma$, respectively.

The number density of swirls, $n_{\rm 2D}$, is computed as the ratio between the average number of identified swirls at each height, $N_{\rm s}$, and the area of a horizontal plane through the simulation domain, $A_{\rm box} = 9.6\,\times\,9.6\,{\rm Mm}^2$. The top panel of Fig.\,\ref{fig:test_stat_z} shows that the number density decreases from $n_{\rm 2D} \sim 4\,{\rm Mm}^{-2}$ in the surface layers of the convection zone ($z=-300\,{\rm km}$), reaching a minimum value of $n_{\rm 2D} \sim 1\,{\rm Mm}^{-2}$ at around $z = 200\,{\rm km}$. Turbulent convection is a natural source of vortices, which explains the large abundance of identified swirls below the average optical surface $\tau_{500} = 1$. Into the upper photosphere and chromosphere, the number density increases again up to $n_{\rm 2D} \sim 5\,{\rm Mm}^{-2}$ around $z=1000\,{\rm km}$. The ratio between the statistical numbers of chromospheric and photospheric swirls corroborates the visual impression we note when comparing Figs.\,\ref{fig:test_sim_photosphere} and \ref{fig:test_sim_chromosphere}. This scenario is also in agreement with three-dimensional renderings of the swirling strength criterion\footnote{The swirling strength criterion, $\lambda$, is a mathematical quantity introduced by \citet[][]{1999JFM...387..353Z}. Similar to the vorticity, it detects local curvature in the flow, but is not biased by the presence of shear flows. For further details, we refer the reader  to \citet[][]{2020A&A...639A.118C}.} shown by \citet[][]{2012A&A...541A..68M} and \citet[][]{2021A&A...649A.121B} in numerical simulations. 
However, the origin of this difference is still not well understood. 

The middle panel of Fig.\,\ref{fig:test_stat_z} shows the distribution of the effective radius of the swirls, $r_{\rm eff}$, computed via Eq.\,(\ref{eq:effective_radius}), as a function of height $z$. The median profile, as well as the $1\sigma$ and $2\sigma$ percentiles, are roughly flat in the surface layers of the convection zone and in the chromosphere, while a slight rise characterizes the low photosphere. The median value is $r_{\rm eff} \sim 50\,{\rm km}$ throughout the surface layers of the convection zone and $r_{\rm eff} \sim 60\,{\rm km}$ in the upper photosphere. The distribution is skewed towards larger values, but $97\,\%$ of the  radii of the identified swirls measure less than $100\,{\rm km}$ and $150\,{\rm km}$ in the subsurface region and in the low chromosphere, respectively. Swirls are therefore statistically larger in the upper layers of the simulated domain. This growth can be explained by the expansion of the plasma ascending into the photosphere caused by the steep decrease in mass density \citep[][]{1997A&A...328..229N}. 

The distribution of the effective rotational period, $P_{\rm eff}$, of the identified swirls is shown in the bottom panel of Fig.\,\ref{fig:test_stat_z}. The effective rotational period of a swirl is computed as
\begin{equation}
    P_{\rm eff} = \frac{4\pi}{|\langle R \rangle_{\rm swirl}|}\,, \label{eq:P_eff}    
\end{equation}
where $\langle R \rangle_{\rm swirl}$ is the average Rortex criterion computed over the swirl area. The median of the distribution reaches its peak of $P_{\rm eff} \sim 140\,{\rm s}$ around $z = 400\,{\rm km}$ with a marked skewness towards larger values, that is, towards slower swirls. This result is compatible with the growth of the typical swirl radius seen in the middle panel, which also reaches its maximum value at around the same height. Indeed, the growth in size caused by the expansion of the photospheric plasma causes the swirls to rotate slower because of the conservation of angular momentum. Moreover, the structure of the distribution is also in agreement with the vertical profiles of average vorticity and swirling strength presented by \citet[][]{2011A&A...533A.126M}, \citet[][]{2020A&A...639A.118C}, and \citet[][]{2021A&A...649A.121B}. 

%
%

\subsubsection{Swirls and magnetic fields}
\label{subsubsec:swirls_and_magnetic_fiels}

\begin{figure}
        \centering
        \resizebox{\hsize}{!}{\includegraphics{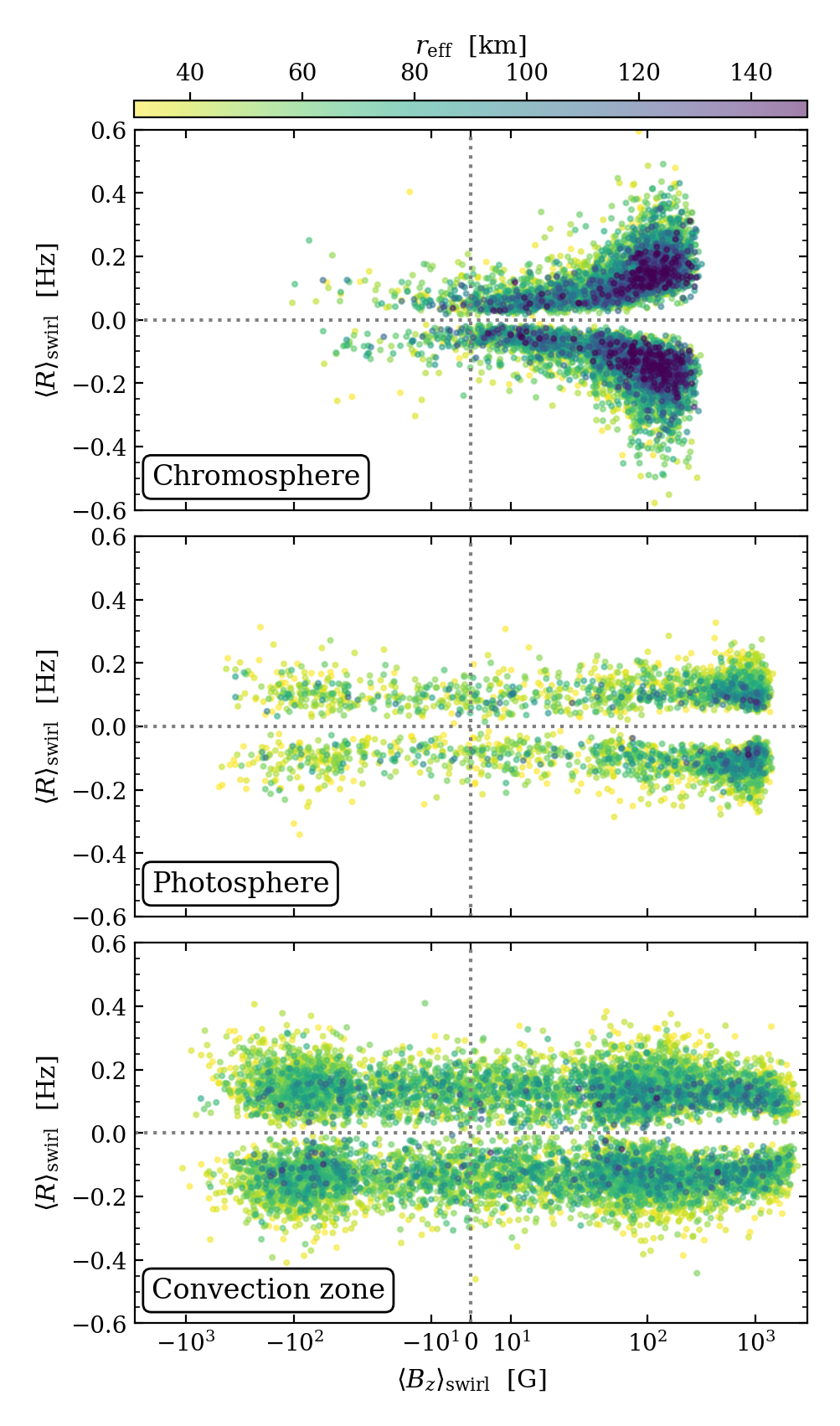}}
        \caption{Bivariate distribution of rotational and magnetic characteristics of vortices at $z=700\,{\rm km}$ (chromosphere, top),  $z=100\,{\rm km}$ (photosphere, middle), and $z=-200\,{\rm km}$ (surface layers of the convection zone, bottom). Every identified vortex in these layers is represented by a scatter point according to the Rortex criterion $\langle R \rangle_{\rm swirl}$ and the vertical magnetic field $\langle B_{z} \rangle_{\rm swirl}$ averaged over their area. The effective radius $r_{\rm eff}$ of the vortex is color coded.}
        \label{fig:test_Stat_Bz_Rz_r}
\end{figure}

Next, we investigate the relation between the vertical magnetic field, $B_z$, and the properties of the identified swirls. Figure \ref{fig:test_Stat_Bz_Rz_r} shows the bivariate distribution of the average Rortex criterion $\langle R \rangle_{\rm swirl}$ and the average vertical magnetic field $\langle B_{z} \rangle_{\rm swirl}$ of the identified swirls. The averages are taken over the area of the swirls in the three horizontal sections corresponding to the surface layers of the convection zone, the photosphere, and the chromosphere. The different panels correspond to the swirls identified at the bottom of the chromosphere ($z=700\,{\rm km}$), in the photosphere ($z=100\,{\rm km}$), and in the surface layers of the convection zone ($z=-200\,{\rm km}$), as shown in Fig.\,\ref{fig:sim_properties}. The effective radius of the swirl is color coded. 

The distributions are symmetric with respect to the sign of the Rortex criterion in all three panels. Therefore, there is no preferred orientation for small-scale swirls in the simulated solar atmosphere. This result is in agreement with the observations reported by, for example, \citet[][]{2018ApJ...869..169G} and \citet[][]{2019NatCo..10.3504L}. 

In the surface layers of the convection zone, swirls are almost homogeneously distributed with respect to the vertical magnetic field for $|\langle B_{z} \rangle_{\rm swirl}| \lesssim 10^2\,{\rm G}$. The properties of these swirls generated by turbulence are expected to be independent of the magnetic field in weakly magnetized regions, because in such circumstances the magnetic field does not affect the dynamics of the plasma. However, an over-density of swirls can be found for $|\langle B_{z} \rangle_{\rm swirl}| \gtrsim 10^2\,{\rm G}$, especially in the positive-polarity end, meaning that swirls tend to be particularly associated with hecto-Gauss magnetic fields. Magnetic flux concentrations can impact the convective dynamics below the solar surface and couple it to the photosphere \citep[see, e.g.,][Fig.\,2]{2021A&A...649A.121B}, 
so that swirls in highly magnetized subsurface regions are part of three-dimensional photospheric vortical structures. We do not find any particular pattern regarding the effective radius of the swirls. The majority of the identified swirls measure between $\sim 30\,{\rm km}$ and $\sim 100\,{\rm km}$, which is in accordance with the results of  Fig.\,\ref{fig:test_stat_z}.  

A clear asymmetry towards positive-polarity magnetic fields is noticeable in the middle and top panels of Fig.\,\ref{fig:test_Stat_Bz_Rz_r}, which correspond to photospheric and chromospheric layers, respectively. We already encountered this asymmetry in the polarity of the vertical magnetic field in Figs.\,\ref{fig:test_sim_photosphere} and \ref{fig:test_sim_chromosphere}, and its origin can be traced back to the initial conditions of the numerical simulation. Indeed, even after relaxation, the polarity of the initial magnetic field persists in most of the photospheric magnetic flux concentrations and within the magnetic canopy of the low chromosphere. The fact that most photospheric and chromospheric vortices are found in regions with positive-polarity magnetic field is therefore a consequence of our choice of the initial condition to mimic a magnetic network patch of preferred polarity; we expect this asymmetry to be lifted in numerical simulations with no preferred initial magnetic configuration. In the surface layers of the convection zone, the initial imbalance is leveled out by the action of a subsurface small-scale turbulent dynamo \citep[see, e.g.,][]{2014ApJ...789..132R}, which ultimately generates the negative-polarity magnetic flux concentrations hosting the less frequent swirls on the left side of the top and middle panels of Fig.\,\ref{fig:test_Stat_Bz_Rz_r}.

On average, chromospheric swirls are larger than photospheric ones. We discuss this difference in Sect.\,\ref{subsubsec:vertical_distribution_of_the_main_properties}. However, we notice that the largest swirls in the photosphere and in the chromosphere are found in correspondence with strong magnetic fields. In Fig.\,\ref{fig:test_sim_3dstructures}, we see that most of the coherent three-dimensional vortical structures in the simulated atmosphere are anchored in photospheric magnetic flux concentrations. Therefore, magnetically dominated regions appear to provide preferable conditions for the creation and preservation of large and coherent vortical structures extending throughout the photosphere and into the chromosphere. 

The distribution in the top panel of Fig.\,\ref{fig:test_Stat_Bz_Rz_r} forms a ``butterfly'' pattern, which hints at the existence of a relation between the Rortex criterion and the vertical magnetic field in chromospheric swirls. The stronger the magnetic field hosting the vortex, the faster it rotates. Moreover, the relation also depends on the radius of the swirl, as the growth of $\langle R \rangle_{\rm swirl}$ as a function of $\langle B_{z} \rangle_{\rm swirl}$ is reduced for larger swirls. In the following, we propose a simple analytical model to explain this relation.

\begin{figure}
        \centering
        \resizebox{\hsize}{!}{\includegraphics{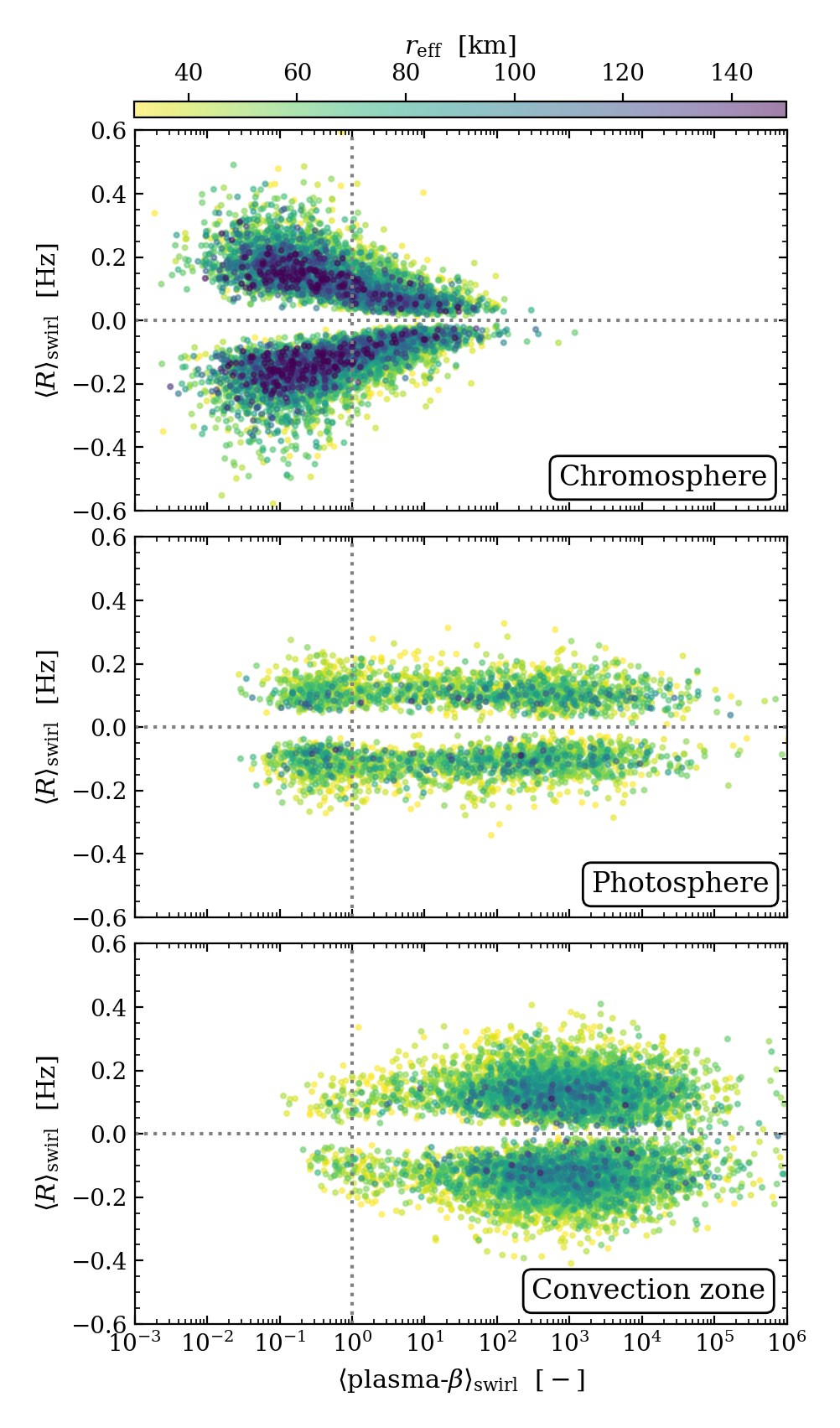}}
        \caption{Bivariate distribution of the Rortex criterion, $\langle R \rangle_{\rm swirl}$, and plasma-$\beta$ averaged over the area of the identified vortices at $z=700\,{\rm km}$ (chromosphere, top), at  $z=100\,{\rm km}$ (photosphere, middle), and at $z=-200\,{\rm km}$ (surface layers of the convection zone, bottom). The effective radius $r_{\rm eff}$ of the vortex is color coded.}
        \label{fig:stat_beta_R_r}
\end{figure}

Figure \ref{fig:stat_beta_R_r} is analogous to Fig.\,\ref{fig:test_Stat_Bz_Rz_r} but shows the average plasma-$\beta$ over the swirl area, $\langle\beta \rangle_{\rm swirl}$, instead of the average vertical magnetic field, $\langle B_z \rangle_{\rm swirl}$. Similar to Fig.\,\ref{fig:test_Stat_Bz_Rz_r}, the distribution is symmetric with respect to the sign of the average Rortex criterion, $\langle R \rangle_{\rm swirl}$, in all panels, showing no preferred direction of rotation for the identified swirls. In the surface layers of the convection zone, the vast majority of swirls are found in $\beta > 1$ conditions, which means that the gas dominates over the magnetic pressure. These swirls are most certainly induced by the turbulent dynamics of the convection zone. 

In the photosphere, we observe the emergence of two different populations of swirls. The first group, characterized by plasma-$\beta > 1$, has the same convective origin as those featured in the bottom panel. A second collection of swirls is instead found in $\beta \lesssim 1$ conditions, where the magnetic field dictates the dynamics of the plasma. The swirls belonging to the second group are embedded in strong magnetic flux concentrations and can represent the footpoints of the coherent three-dimensional structures observable in Fig.\,\ref{fig:test_sim_3dstructures}. 

Finally, in the top panel of Fig.\,\ref{fig:stat_beta_R_r}, we notice a butterfly pattern similar to that seen in Fig.\,\ref{fig:test_Stat_Bz_Rz_r}. The fastest and largest swirls are characterized by low $\beta$ values, which correspond to the ones with high $\langle B_z \rangle_{\rm swirl}$ in Fig.\,\ref{fig:test_Stat_Bz_Rz_r}. The dynamics of these swirls are dominated by the magnetic field and the model proposed in Sect.\,\ref{subsubsec:models_of_chromospheric_swirls} qualitatively applies. 

There are also a large number of chromospheric swirls for which the local $\beta$ is larger than one. These swirls populate weakly magnetized areas of the chromosphere. In these regions, purely hydrodynamical mechanisms, such as baroclynic forces or shocks, could be at the origin of these chromospheric swirls, which appear to be locally produced.

%
%

\subsubsection{A simple model of magnetic chromospheric swirls}
\label{subsubsec:models_of_chromospheric_swirls}

\begin{figure}
        \centering
        \resizebox{\hsize}{!}{\includegraphics{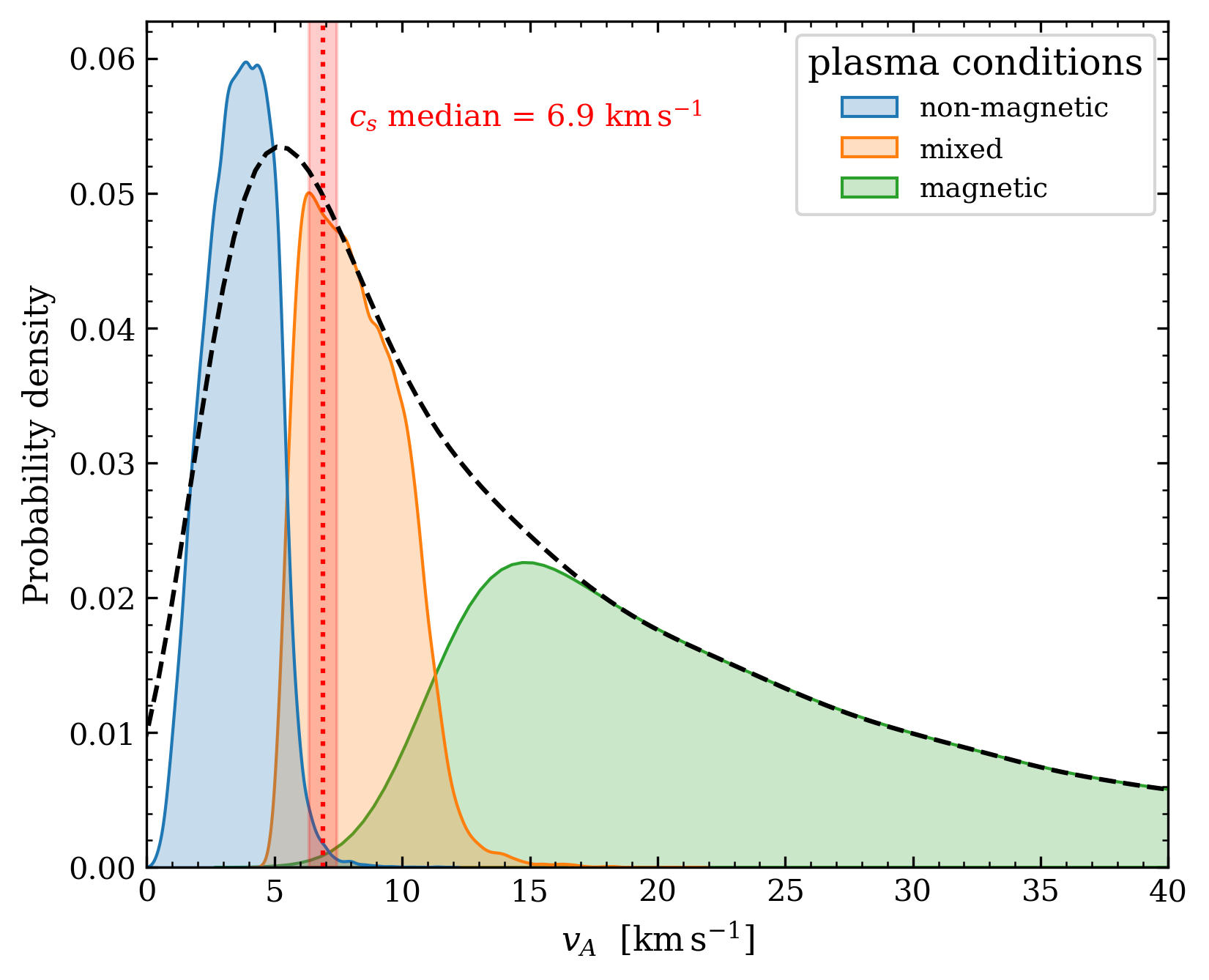}}
        \caption{Probability distributions of the average Alfvén speed, $v_{\rm A}$, computed over the effective area of swirls identified between $z=600\,{\rm km}$ and $z=900\,{\rm km}$. The total kernel density estimate is shown in black. The distributions are divided according to the average plasma-$\beta$ conditions computed over the effective area of the  swirls. The nonmagnetic category corresponds to plasma-$\beta > 2.0$, the mixed category corresponds to $0.5 <$ plasma-$\beta \leq 2.0$, and the magnetic category corresponds to plasma-$\beta \leq 0.5$.  The red area and vertical dashed line correspond to the 5-95 percentile range and the median of the average sound speed computed over the effective area of the  swirls, respectively.}
        \label{fig:estimate_rotationa_speed}
\end{figure}
\begin{figure}
        \centering
        \resizebox{\hsize}{!}{\includegraphics{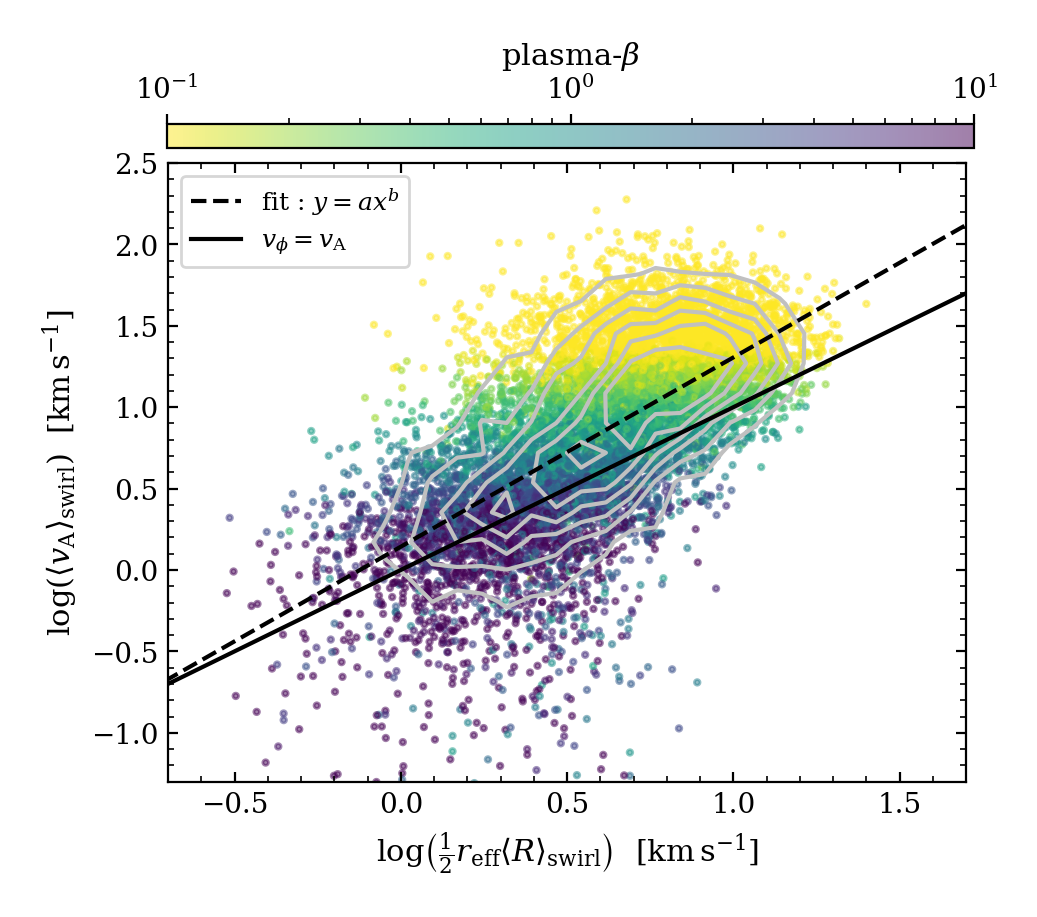}}
        \caption{Bivariate distribution of the local average Alfvén speed, $\langle v_{\rm A} \rangle_{\rm swirl}$, and the average estimated rotational velocity of the swirl, $\langle v_{\phi} \rangle_{\rm swirl} = \frac{1}{2}r_{\rm eff} \langle R \rangle_{\rm swirl}$ at the bottom of the chromosphere ($z=700\,{\rm km}$). The averages are computed over the swirl area. The average plasma-$\beta$ over the swirl area is color coded. A power-law fit of the type $y = a x^b$ is shown in dashed black. The fitted parameters are $a = 0.16,\, b = 1.13$. Density contours of the scattered points are shown in gray.}
        \label{fig:Test_Stat_BrR_relation}
\end{figure}

Let us consider a chromospheric swirl coupled to a strong magnetic flux tube, such as those identified above. For simplicity, we assume their shape to be a cylinder. We further assume the system to be in stationary magnetohydrodynamic radial equilibrium and to rotate as a rigid body. The MHD momentum equation in the radial coordinate, $r$, can then be written as
\begin{equation}
    \partial_r p - \left( \boldsymbol{B}\cdot\boldsymbol{\nabla} \right)B_r = \rho \frac{v_{\phi}^2}{r}\,,\label{eq:MHD_mom_eq_swirl}
\end{equation}
where $p = p_{\rm g} + p_{\rm m}$ is the total pressure, that is the sum of the gas pressure, $p_{\rm g}$, and the magnetic pressure, $p_{\rm m} =  \boldsymbol{B}^2/8\pi$; $\boldsymbol{B} = (B_r, B_{\phi}, B_z)$ is the magnetic field in cylindrical coordinates; $\rho$ is the plasma density; and $v_{\phi}$ is the plasma rotational velocity. In the chromosphere, a strong magnetic flux tube is characterized by plasma-$\beta \lesssim 1$, and therefore it is safe to further assume that the total pressure $p$ is dominated by the magnetic field component, that is, $p \simeq p_{\rm m}$. Moreover, the rigid-body rotational velocity of the plasma is related to the angular velocity by $v_{\phi} = \Omega r$. Taking this into account, Eq.\,(\ref{eq:MHD_mom_eq_swirl}) becomes
\begin{equation}
    \frac{1}{8\pi}\partial_r\left( \boldsymbol{B}^2 \right)  - \left( \boldsymbol{B}\cdot\boldsymbol{\nabla} \right)B_r = \rho \Omega^2 r\,.\label{eq:MHD_mom_eq_swirl_2}
\end{equation}

We model a chromospheric section of the magnetic flux tube with purely vertical magnetic field $B_z = B(r)$ and density $\rho$. Therefore, the magnetic field inside the cylinder is $\boldsymbol{B} = B(r) \boldsymbol{e}_z$. In this scenario, Eq.\,(\ref{eq:MHD_mom_eq_swirl_2}) can be further simplified into
\begin{equation}
    \frac{1}{8\pi} \partial_r \left(B^2\right) = \rho \Omega^2 r\,, \nonumber
\end{equation}
and integrated over the radius of the swirl $r$, leading to
\begin{equation}
    \frac{1}{4\pi \rho } \left(B(r)^2 - B(0)^2\right) = \Omega^2 r^2\,. \label{eq:MHD_mom_eq_swirl_3}
\end{equation}

A physical solution to the equation above exists only if $B(r)^2 > B(0)^2$, that is, if the rotation of the plasma is supported by a negative magnetic pressure gradient toward the vortex core. Indeed,   the cores of vortices within large magnetized regions are often found to be associated with reduced magnetic pressure. We provide an example of such an event in Appendix \ref{app:A}.

We recognize the local Alfvén speed, $v_{\rm A}(r) = B(r)/\sqrt{4\pi\rho}$, and the rotational velocity of the swirl, $v_{\phi} = \Omega r$, on the left- and right-hand sides of Eq.\,(\ref{eq:MHD_mom_eq_swirl_3}), respectively,
\begin{equation}
    v_{\rm A}(r)^2 - v_{\rm A}(0)^2 = v_{\phi}^2\,, \label{eq:model_relation}
\end{equation}
where $v_{\rm A}(0)$ is the Alfvén speed computed in the vortex center. The above equation states that the Alfvén speed in the magnetic flux tube is the upper limit to the swirl rotational velocity. If we assume the magnetic field in the vortex core to be weak enough compared to the bulk of the flux tube, then the swirl rotates at approximately the local Alfvén speed, $v_{\phi} \approx v_{\rm A}$. 

For $r=0$, Eq.\,(\ref{eq:model_relation}) predicts $v_{\phi} = 0$, which is consistent with the assumption of rigid-body rotation. Moreover, in Appendix \ref{app:A}, we show that the structure of the chromospheric swirl shown in Fig.\,\ref{fig:test_sim_vortexB} qualitatively agrees with the model presented above.

For a consistency test of Eq.\,(\ref{eq:model_relation}), we estimate the expected rotational velocity, $v_{\phi}^{\rm exp}$, of a chromospheric swirl and compare it to the Alfvén speed in chromospheric swirls of low plasma-$\beta$ conditions. We base our estimation on typical values of the effective radius, $r_{\rm eff}$, and the Rortex criterion, $R$, for chromospheric swirls in plasma-$\beta \ll 1$. From Fig.\,\ref{fig:stat_beta_R_r}, we infer that such values are $r_{\rm eff} \sim 120\,{\rm km}$ and $R \sim 0.2\,{\rm Hz}$.
    
Using the formula

\begin{equation}
v_{\phi} = \Omega r \approx \frac{R}{2}  r_{\rm eff} \, , \label{eq:estimate_rotational_speed}
\end{equation}

we calculate an expected rotational velocity for swirls at $z=700\,{\rm}$ in low plasma-$\beta$ conditions of $v_{\phi}^{\rm exp} \sim 12\,{\rm km}\,{\rm s}^{-1}$. 

Figure\,\ref{fig:estimate_rotationa_speed} shows the probability distribution of the average Alfvén speeds computed over the effective area of swirls identified in the low chromosphere ($600\,{\rm km} < z < 900\,{\rm km}$). There, we differentiate between swirls in magnetic conditions (plasma-$\beta \leq 0.5$), mixed conditions ($0.5 <$  plasma-$\beta \leq 2.0$), and nonmagnetic conditions (plasma-$beta > 2.0$), for a succinct labeling. The expected rotational velocity of chromospheric swirls inferred from Fig.\,\ref{fig:stat_beta_R_r} and Eq.\,(\ref{eq:estimate_rotational_speed}) and the distribution of Alfvén speeds averaged over the effective area of swirls in low plasma-$\beta$ conditions are consistent with the analytical model and Eq.\,(\ref{eq:model_relation}). For comparison, the distribution of the average sound speeds computed over the effective area of the  swirls is outlined by its median together with the $5-95$ percentile range. The estimated rotational speed and the rotational speed derived from the model are clearly above the sound speed.

Figure \ref{fig:Test_Stat_BrR_relation} shows a scatter plot of all the identified vortices at $z=700\,{\rm km}$ as a function of the Alfvén speed $\langle v_{\rm A} \rangle_{\rm swirl}$ and the estimated rotational velocity $\langle v_{\phi} \rangle_{\rm swirl} = \frac{1}{2}r_{\rm eff}\langle R \rangle_{\rm swirl}$, both averaged over the swirl area. A large dispersion characterizes the distribution, which is expected given the rough approximations made in deriving Eq.\,(\ref{eq:model_relation}), but a linear trend is perceivable. The solid black line shows the relation $v_{\phi} = v_{\rm A}$. For $\beta \lesssim 1$, we see that $v_{\phi} \lesssim v_{\rm A}$ for almost all swirls, confirming that $v_{\rm A}$ is an upper limit for $v_{\phi}$. However, this limit applies less well for weak field swirls with $\beta \gtrsim 1$.

We fitted a power-law function of the type $y = a x^b$ to the data. The resulting curve is represented by the black dashed line in the log--log plot of Fig.\,\ref{fig:Test_Stat_BrR_relation}. The fitted exponent is $b=1.16$, which is quite close to the modeled linear exponent $b=1$. Another measure of the linear correlation between $\langle v_{\rm A} \rangle_{\rm swirl}$ and $r_{\rm eff} \langle R \rangle_{\rm swirl}$ for the identified chromospheric swirls can be obtained in the form of the Pearson's correlation coefficient $r_{\rm P}$. For the dataset shown in Fig.\,\ref{fig:Test_Stat_BrR_relation}, we obtain $r_{\rm P} = 0.45$, which demonstrates a discrete degree of linear correlation between these two quantities. 

\subsubsection{Torsional Alfvénic waves}
\label{subsubsec:alfvenic_properties}

\begin{figure}
        \centering
        \resizebox{\hsize}{!}{\includegraphics{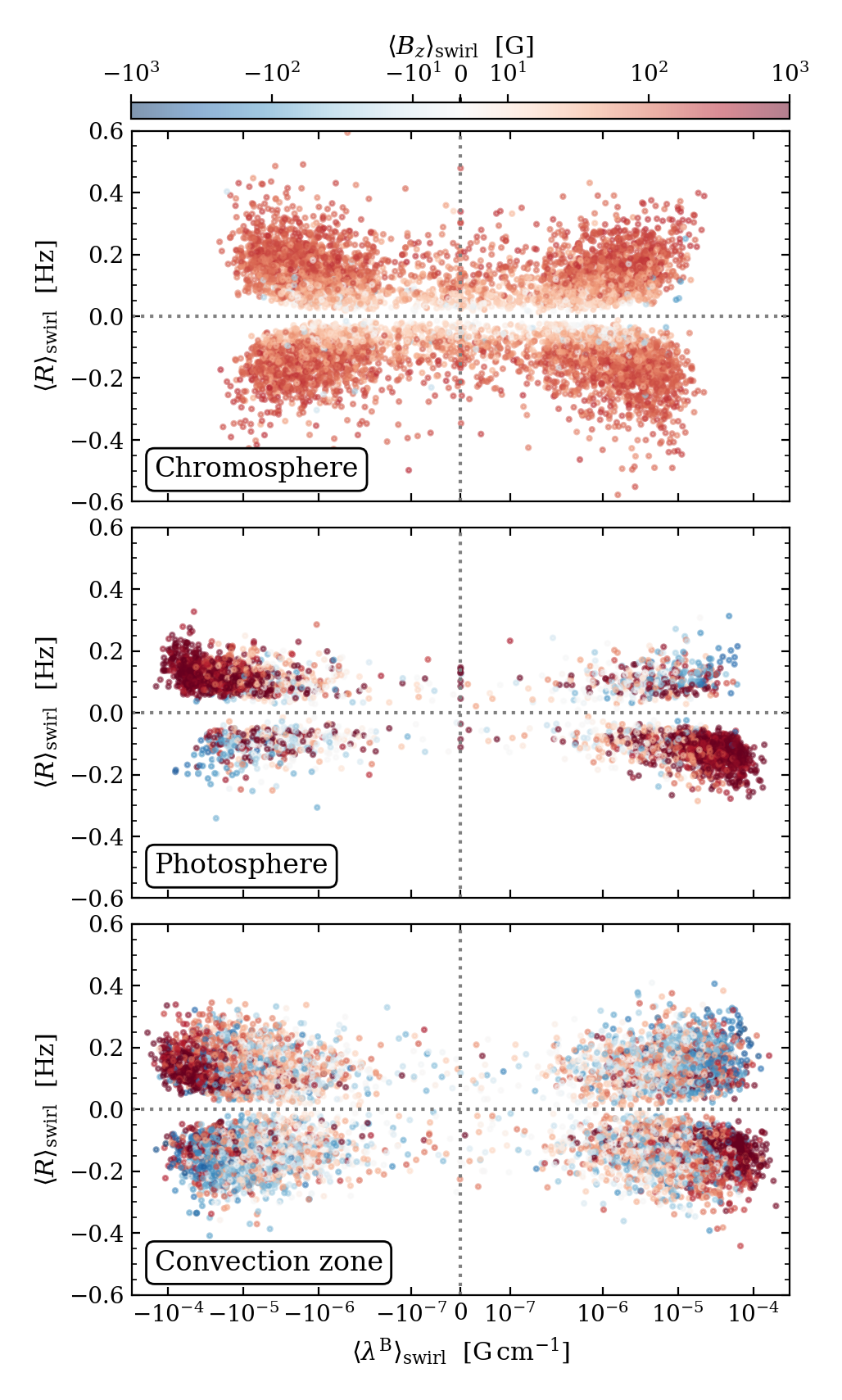}}
        \caption{Bivariate distribution of the rotational characteristics of vortices at $z=700\,{\rm km}$ (chromosphere, top),  $z=100\,{\rm km}$ (photosphere, middle), and $z=-200\,{\rm km}$ (surface layers of the convection zone, bottom). Every identified vortex in these layers is represented by a scatter point according to the Rortex criterion $\langle R \rangle_{\rm swirl}$ and the magnetic swirling strength criterion $\langle \lambda^{\rm B} \rangle_{\rm swirl}$ averaged over the swirl area. The vertical magnetic field $\langle B_z \rangle_{\rm swirl}$ averaged over the swirl area is color coded.}
        \label{fig:test_Stat_Msz_Rz_Bz}
\end{figure}

We also investigated the correlation between swirls in the simulated solar atmosphere and perturbations in the magnetic field lines. \citet[][]{2021A&A...649A.121B} 
reported that a toroidal perturbation in the predominantly vertically directed magnetic field can be found in upwardly propagating pulses of swirling plasma. The same authors introduced the magnetic swirling strength, $\lambda^{\rm B}$, which is a measure for the toroidal components, or twists, in magnetic flux tubes. The simultaneous presence of a twist in the magnetic field lines and a vortical motion in the plasma may hint at the presence of torsional Alfvénic waves propagating cojointly with the rotating magnetic flux concentration.  

Figure \ref{fig:test_Stat_Msz_Rz_Bz} shows the bivariate distribution of the swirls identified at the bottom of the chromosphere ($z = 700\,{\rm km}$), in the photosphere ($z = 100\,{\rm km}$), and in the surface layers of the convection zone ($z = -200\,{\rm km}$) as a function of the Rortex, $\langle R \rangle_{\rm swirl}$, and magnetic swirling strength, $\langle \lambda^{\rm B} \rangle_{\rm swirl}$, averaged over the swirl area. The average strength of the vertical magnetic field over the swirl area, $\langle B_z \rangle_{\rm swirl}$, is color coded. 

If we do not consider the polarity of the vertical magnetic field, the swirls identified in the surface layers of the convection zone (bottom panel) are distributed almost symmetrically with respect to $\langle R \rangle_{\rm swirl}$ and $\langle \lambda^{\rm B} \rangle_{\rm swirl}$. Once more, we explain this symmetry with the isotropical turbulence that dominates the surface layers of the convection zone. However, the magnetic field orientation reveals a pattern that is even more prominent in the photosphere (middle panel): most of the swirls embedded in positive-polarity magnetic fluxes are concentrated in the top-left and bottom-right quadrants of the bivariate distribution, while the ones associated with negative-polarity magnetic fields are found in the top-right and bottom-left quadrants. 

The excess of red points ($\langle B_z \rangle_{\rm swirl} > 0\,{\rm G}$) in the photospheric and chromospheric distributions is due to the initial conditions of the present simulations. Moreover, the pattern is less pronounced in the surface layers of the convection zone because of the swirls that are randomly generated by turbulence and are not part of coherent photospherical structures.

The pattern revealed by the middle and bottom panels of Fig.\,\ref{fig:test_Stat_Msz_Rz_Bz} can be explained if we consider the swirls to be Alfvénic in nature, as proposed by \citet[][]{2019NatCo..10.3504L} and \citet[][]{2021A&A...649A.121B}. 
For a vertical magnetic field $\boldsymbol{B} = B_z \boldsymbol{e_z}$ and an incompressible plasma in magneto-hydrostatic equilibrium in the ideal MHD approximation, a torsional Alfvén wave is characterized by velocity and magnetic field perturbations, $\boldsymbol{v}$ and $\boldsymbol{b}$, that obey \citep[see, e.g.,][Chap.\,4]{2014masu.book.....P} 

\begin{equation}
    \boldsymbol{v} = - \frac{\omega \boldsymbol{b}}{\boldsymbol{k}\cdot\boldsymbol{B}}\,,\label{eq:alfvenwave}
\end{equation}
where $\omega$ is the angular frequency of the plane wave and $\boldsymbol{k}$ is the wave-vector indicating the propagation direction. For a vertically propagating torsional Alfvén wave, that is, $\boldsymbol{k} = k \boldsymbol{e}_z$ with $k>0$, Eq.\,(\ref{eq:alfvenwave}) can be simplified as
\begin{equation}
    \boldsymbol{v} = - \frac{v_{\rm A}}{B_z} \boldsymbol{b}\,, \label{eq:alfvenwave_simple}
\end{equation}
where $v_{\rm A} > 0$ is the local Alfvén speed, while $\boldsymbol{v}$ and $\boldsymbol{b}$ are perturbations in the horizontal plane. 

From Eq.\,(\ref{eq:alfvenwave_simple}) we conclude that the perturbations $\boldsymbol{v}$ and $\boldsymbol{b}$ are parallel or anti-parallel depending on the polarity of the vertical magnetic field, that is, on the sign of $B_z$. If we use the Rortex and the magnetic swirling strength criteria as proxies to quantify such perturbations, then we can write
\begin{equation}
    {\rm sign}{\left( R \lambda^{\rm B} \right)} = - {\rm sign}{\left(B_z\right)}\,, \label{eq:alfven_sign}
\end{equation}
and the distributions in the surface layers of the convection zone and photosphere of Fig.\,\ref{fig:test_Stat_Msz_Rz_Bz} appear to statistically follow this relation. The clockwise vortex associated with the counterclockwise twist of the positive-polarity magnetic field lines shown in Fig.\,\ref{fig:test_sim_vortexB} is a practical illustration of Eq.\,(\ref{eq:alfven_sign}).

The Alfvénic pattern encountered in the lower and middle layers of the simulation box seems to disappear in the chromosphere (top panel of Fig.\,\ref{fig:test_Stat_Msz_Rz_Bz}).
A high degree of symmetry is restored in the distribution, although the polarity of the magnetic field is predominantly positive. If the chromospheric swirls were associated with upwardly propagating Alfvén waves, we would expect Eq.\,(\ref{eq:alfven_sign}) to be respected and the scatter points to populate mainly the top-left and bottom-right quadrants of the plot. 

We find two explanations for the systematic violation of Eq.\,(\ref{eq:alfven_sign}) in the simulated chromosphere. First, in Sect.\,\ref{subsubsec:vertical_distribution_of_the_main_properties}, we show that chromospheric swirls are more abundant than photospheric ones. Therefore, a large fraction of swirls must be generated locally in the simulated chromosphere and, as they are not linked to a photospheric coherent structure, they do not share the same properties. 
Second, the boundary conditions of the simulation force the magnetic field to be strictly vertical at the top boundary, which may cause upwardly directed Alfvénic waves to be reflected. In that case, the wave-vector becomes $\boldsymbol{k} = -k \boldsymbol{e}_z$, Eq.\,(\ref{eq:alfvenwave_simple}) picks up a minus sign on the right-hand side, and therefore Alfvénic waves propagating downward are characterized by parallel perturbations $\boldsymbol{v}$ and $\boldsymbol{b}$ when embedded in positive-polarity magnetic fields. Consequently, swirls associated with downwardly directed torsional Alfvén pulses would populate the top-right and bottom-left quadrants of the top panel of Fig.\,\ref{fig:test_Stat_Msz_Rz_Bz}. 
We note that the narrower distribution of data points over $\langle \lambda^{B}\rangle_{\rm swirl}$ in the chromosphere compared to the photosphere and surface layers of the convection zone is due to the fact that the magnetic swirling strength is proportional to the magnetic field strength, which in turn is much smaller in the chromosphere; it does not indicate a smaller twist angle.

\begin{figure}
        \centering
        \resizebox{\hsize}{!}{\includegraphics{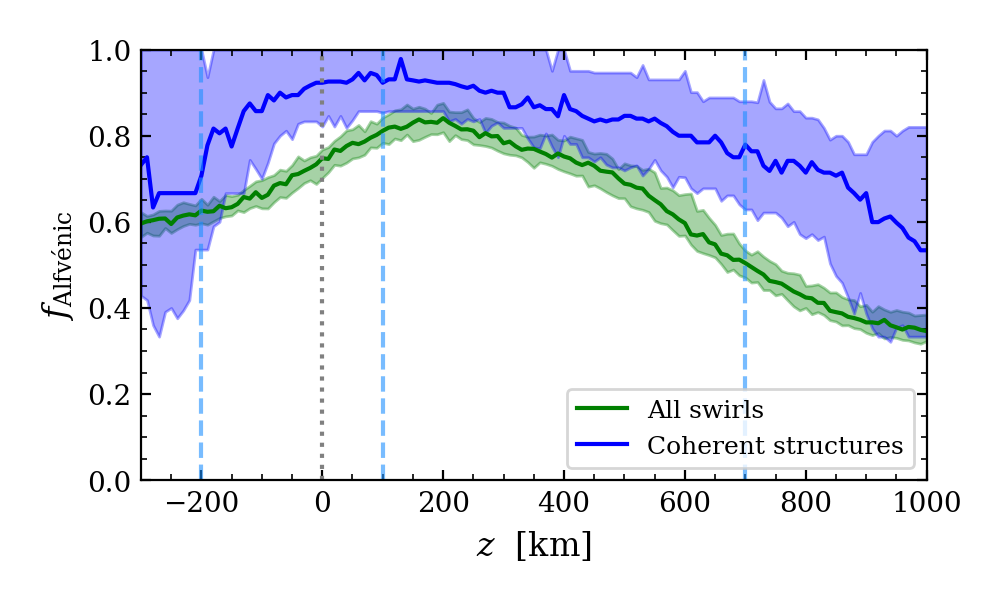}}
        \caption{Fraction of swirls obeying Eq.\,(\ref{eq:alfven_sign}), $f_{\rm Alfvénic}$, as a function of height $z$. The green profile refers to all identified swirls, while the blue curve takes into account only those swirls that form coherent structures connecting the surface layers ($z=0\,{\rm km}$) to the low chromosphere ($z=700\,{\rm km}$). Shaded areas represent statistical standard deviations. The average optical surface $\tau_{500} = 1$ ($z=0\,{\rm km}$) is marked by a dotted line, while the heights of the surface layers of the convection zone, photosphere, and low chromosphere used in the analysis are indicated by dashed blue lines. }
        \label{fig:test_Stat_AlfvenImprints}
\end{figure}

To characterize the abundance of swirls exhibiting imprints of upwardly propagating Alfvénic waves, we computed the fraction of them, $f_{\rm Alfvénic}$, for which Eq.\,(\ref{eq:alfven_sign}) is respected. Figure \ref{fig:test_Stat_AlfvenImprints} shows the results obtained by considering all the swirls identified in any time instances of the simulation (green curve) and those only forming three-dimensional swirling structures (blue curve). In the latter case, only the structures that reach both the surface layers ($z=0\,{\rm km}$) and the low chromosphere ($z=700\,{\rm km}$) are taken into account. 

In the photosphere, approximately $80\,\%$ of all the swirls show perturbations in the plasma and in the magnetic field that are compatible with torsional Alfvénic waves, which is in accordance with the pattern observed in Fig.\,\ref{fig:test_Stat_Msz_Rz_Bz}. However, this fraction decreases as we move upward in the simulation box and falls below $50\,\%$ in the chromosphere. Regarding swirls that belong to coherent three-dimensional structures, we notice a higher fraction at all heights, reaching $\sim90\,\%$ in the photosphere and $\sim80\,\%$ at $z=700\,{\rm km}$. 

Therefore, Fig.\,\ref{fig:test_Stat_AlfvenImprints} suggests that a significant fraction of the identified coherent three-dimensional swirls present characteristics compatible with torsional Alfvénic pulses propagating upward in the simulated solar atmosphere. On the other hand, vortical structures that do not couple the photosphere to the chromosphere appear to show fewer imprints of these waves, and are therefore probably  of a different nature and likely have a different origin.  

%
%

\section{Summary and conclusions}
\label{sec:conclusions}

In this paper, we employed the recently developed SWIRL algorithm to investigate small-scale swirls in radiative MHD numerical simulations of the solar atmosphere. The methodology at the core of this algorithm considers both the local and global properties of the velocity field in the detection process. Therefore, the SWIRL algorithm is specifically tailored to identifying coherent vortical structures, whereas conventional methods, such as the vorticity or the swirling strength, can only recognize local curvatures in the flow. The identification process is automatized through the implementation of a state-of-the-art clustering algorithm. This approach requires minimal interaction with the user, which reduces the risk of human bias in the identification process, and ensures a high level of precision and consistency. In \paperIt,  we validated the robustness of the SWIRL algorithm against noise and turbulence.

In a first stage of the present paper, we tested the reliability of the code in identifying swirls that emerge self-consistently from the simulated photospheric and chromospheric flows. The interplay between magnetic fields, turbulence, convective flows, and shocks significantly increases the complexity of the flow with respect to the tests carried out in \paperIt. In addition, fine-tuning of the algorithm parameters is necessary, especially for the number of stencils, the ``noise'' parameter, and the ``kink'' parameter. We provide the list of parameter values used in this study in Table\,\ref{tab:swirl_params}; we consider these to be suitable default values for applying the SWIRL algorithm to numerical simulations of the solar atmosphere.

The algorithm detected photospheric and chromospheric swirls with high accuracy and precision based on the instantaneous streamlines of the horizontal component of the velocity field. Occasional misidentifications can occur, as shown in Figs.\,\ref{fig:test_sim_photosphere_zoomin} and \ref{fig:test_sim_chromosphere_zoomin}. Moreover, the identification method implemented in the SWIRL algorithm is not Galilean invariant and therefore swirls that are advected at speeds comparable to their rotational velocity could be missed. This shortcoming should not affect photospheric swirls, which are predominantly rooted in intergranular lanes and tightly coupled to magnetic flux concentrations, but it could be relevant to swirls in a dynamical environment such as the chromosphere. Further investigation of this aspect is required in order to improve the performance of the algorithm and to reduce such inaccuracies.
Together with the tests carried out in \paperIt, we conclude that the present SWIRL algorithm is a reliable tool for the identification of swirls in the solar atmosphere and astrophysical flows in general. 

The SWIRL algorithm is currently limited to the identification of vortices in two-dimensional planes. Therefore, in order to investigate the presence of coherent three-dimensional vortical structures extending vertically in the simulation domain, we ran the algorithm on all the horizontal sections of a particular time instance of the CO5BOLD simulation. Successively, we stacked up the identified vortices that were approximately vertically aligned in order to reconstruct the three-dimensional structures. Figure \ref{fig:test_sim_3dstructures} shows an example result of this procedure and demonstrates that the vast majority of small-scale swirls that reach the chromosphere stem from photospheric magnetic flux concentrations. Depending on the intensity and complexity of these magnetic regions, we can observe isolated vortex tubes or multiple swirls that coexist and interact within a magnetic element \citep[see also][]{2021A&A...649A.121B}. 

The procedure adopted in this paper to find three-dimensional swirls is relatively basic and vortical structures in the simulation domain may have been missed. Moreover, this method can outline vertically extending structures only, while horizontally directed vortex tubes and arches have been shown to populate the solar atmosphere as well. An identification code that could handle the three-dimensional flow of a simulation domain (or subdomain) would, in principle, be required to properly characterize and study these structures. As we argue in \paperIt, the SWIRL methodology and algorithm can, in principle, be extended to three dimensions. However, the computational costs that such an upgrade would entail complicate matters considerably, especially regarding the automated clustering task. The identification process on a $960\,\times\,960$ plane of the CO5BOLD simulations typically takes around $\sim 2\,{\rm min}$ on a single CPU. Of this time, approximately $90\,\%$ of the computational time is dedicated to the clustering step. As the size of the data set increases, the percentage of time spent on clustering is expected to rise due to the inherent computational complexity of the clustering algorithm. Consequently, the overall time required for the complete identification process will also increase.

In the second part of this paper, we present a statistical analysis of the properties of small-scale swirls in numerical simulations of the solar atmosphere and near-surface convection zone. Our study indicates that, statistically, around one small-scale swirl can be found in each ${\rm Mm}^2$ of the photosphere, while in the low chromosphere the number density of swirls grows to approximately $4\,{\rm Mm}^{-2}$. Because of these different abundances, approximately three out of four chromospheric swirls must be generated locally in the chromosphere, but the physical mechanism responsible for this generation is still unknown. 

\citet[][]{2012ASPC..463..107S} and \citet[][]{2020A&A...639A.118C} 
presented an analysis of the generation of vortical motions based on the evolution equation of the vorticity and of the swirling strength, respectively. Both papers concluded that the origin of swirling motions in the chromosphere must be traced to the action of magnetic fields. However, in Sect.\,\ref{subsubsec:swirls_and_magnetic_fiels} we show that a considerable number of chromospheric swirls are found in high plasma-$\beta$ regions, which indicates that hydrodynamical forces or shocks may be responsible for part of the small-scale swirls in the upper solar atmosphere.  

If we extrapolate the obtained number densities to the whole Sun, our results hint at the steady presence of $\sim 6\times10^6$ and $\sim 2\times10^7$ swirls in the photosphere and the chromosphere, respectively. These numbers greatly exceed previous estimations based on simulations and observations reported in the literature \citep[see, e.g.,][]{2012Natur.486..505W, 2017A&A...601A.135K, 2018ApJ...869..169G, 2019ApJ...872...22L, 2019NatCo..10.3504L, 2022A&A...663A..94D}. On the other hand, our analysis reveals that the average size of the swirls in the simulated atmosphere settles down to around $50-60\,{\rm km}$ in radius, although larger vortices can be systematically found in the chromosphere. In summary, swirls may be more numerous and smaller than previously thought.

For comparison, we mention in the following a few studies addressing the number densities and typical radii of small-scale swirls in the solar atmosphere. A more comprehensive list of these values can be found in \citet[][]{2022...ISSI}. \citet[][]{2012Natur.486..505W} 
counted on average $\sim 2.0\times10^{-3}\,{\rm Mm}^{-2}$ ($3.8\,{\rm arcmin}^{-2}$) long-lived chromospheric swirls with a typical radius of  $1.4\times10^3$\,{\rm km} ($2.0\,{\rm arcsec}$) from observations obtained with the CRisp Imaging SpectroPolarimeter (CRISP) instrument of the Swedish 1m Solar Telescope (SST). Automated surveys have been carried out by \citet[][]{2018ApJ...869..169G} and \citet[][]{2019ApJ...872...22L} 
on photospheric observations obtained with CRISP/SST and with the Solar Optical Telescope (SOT) on board the Hinode satellite, respectively. The first study identified on average $2.7\times10^{-2}\,{\rm Mm}^{-2}$ swirls with a mean radius of $290\,{\rm km}$, while the second one found number densities that are closer to our result, namely $2.4\times10^{-1}\,{\rm Mm}^{-2}$, but with an average radius of $280\,{\rm km}$. Using a new automated identification method based on the morphological characterization of ${\rm H}\alpha$ spectral lines \citep[][]{2021SoPh..296...17D}, \citet[][]{2022A&A...663A..94D} 
found a number density of chromospheric swirls of $8\times10^{-2}\,{\rm Mm}^{-2}$ and an average radius of $1.3\times10^3\,{\rm km}$ from CRISP/SST observations.
From numerical simulations, \citet[][]{2017A&A...601A.135K} 
also detected a relatively high number of chromospheric swirls with an average number density of $8.6\times10^{-1}\,{\rm Mm}^{-2}$. However, in this case, the average radius of the identified swirls was $338\,{\rm km}$.

Nevertheless, we would not recommend a blunt comparison between the results presented in this paper and previous results found in the literature. First, the identifications performed on observational data heavily rely on the methods used to estimate the horizontal velocity fields. For example, LCT techniques should be used with caution as they present several limitations, especially in estimating granular and subgranular flows \citep[][]{2013A&A...555A.136V, 2018SoPh..293...57T}. 
To our knowledge, the only study that is not affected by this issue is the one presented by \citet[][]{2022A&A...663A..94D}, because these authors detected swirls directly from chromospheric filtergrams. One possible solution to consider is a deep learning approach, as proposed by \citet{2017A&A...604A..11A}. However, it is important to be aware that the simulations on which the models are trained may introduce bias into the results if their vortical flows are not consistent with the real ones.

Second, the properties of vortical motions appear to be heavily dependent on the spatial resolution available, as shown by \citet[][]{2020ApJ...894L..17Y} 
in numerical simulations. Other details regarding the simulations, such as the initial and boundary conditions or the strength of the magnetic field, can also deeply affect the characteristics of vortical motions (see, e.g., Appendix A of \citet{2021A&A...649A.121B} 
or \citet{2022csss.confE...4C} 
for simulations with different initial magnetic fields)
A comprehensive investigation of the influence of the numerical setup on the characteristics of vortices is an essential step towards a deeper understanding of their formation and evolution in numerical simulations of the solar atmosphere.

Finally, different datasets and different automated algorithms have been used for the identification of swirls in the solar atmosphere. A comparative study between the available algorithms would be necessary to assess strengths and weaknesses of the different detection methods. 

Given the clear correlation between magnetic flux concentrations and vortical motions, we investigated how the properties of the small-scale swirls vary as a function of the vertical component of the magnetic field. We find indications of a relation between the vertical magnetic field, the angular velocity, and the size of chromospheric swirls. We explain this relation with a simple model of a homogeneously dense magnetic flux tube in a low-plasma-$\beta$ environment with a magnetic pressure gradient that supports its rotation. 

This model assumes stationary radial equilibrium and rigid-body rotation. We acknowledge that these assumptions are only very basic and do not accurately capture the complex nature of chromospheric swirls. For example, swirls do not rigidly rotate \citep[see, e.g.,][]{2020ApJ...898..137S} and flows in the highly dynamical chromosphere are not stationary. Nevertheless, they allow a straightforward analytical analysis and interpretation of the swirl properties and of our statistical results. The model suggests that chromospheric swirls can rotate at maximum speeds that approach the local Alfvén speed, and the data gathered from the simulation support this conclusion. To our knowledge, this result represents a new property of chromospheric swirls that has not yet been investigated, and that could have profound implications for the total energy transport associated with small-scale vortical motions in the solar atmosphere.

\citet[][]{2016A&A...586A..25P} 
observed a chromospheric swirl measuring $0.5-1.0\,{\rm Mm}$ and rotating at an average speed of $13\,{\rm km}\,{\rm s}^{-1}$ with CRISP/SST. Although there is no available information regarding the magnetic field for this particular event, the observed rotational velocity is in the range of typical Alfvén speeds in chromospheric conditions. Other observational studies, such as those of \citet[][]{2009A&A...507L...9W}, \citet[][]{2013ApJ...768...17M}, \citet[][]{2019NatCo..10.3504L}, and \citet[][]{2020A&A...639A..59M},  suggest slower average rotational velocities for chromospheric swirls, rarely exceeding $2\,{\rm km}\,{\rm s}^{-1}$. We expect future high-resolution observations with the Daniel K. Inouye Solar Telescope (DKIST) to shed light on the real rotational velocities of swirls in the solar atmosphere.

Finally, we carried out a statistical analysis in order to investigate the possible Alfvénic nature of photospheric and chromospheric swirls. We find a clear relation between the orientation of the identified swirls, the orientation of the toroidal magnetic perturbations in the swirling area, and the polarity of the vertical magnetic field emerging from the data, in particular in the photosphere. In $80\,\%$ of the identified photospheric swirls, this relation is compatible with the propagation of torsional Alfvén waves according to Eq.\,(\ref{eq:alfvenwave_simple}). In the chromosphere, the correlation between swirls and Alfvénic waves seems to vanish, probably because of the local generation of vortical motions, which do not have a photospheric counterpart. However, when considering only those swirls that extend from the photosphere to the chromosphere, we find 90\% and 80\% of them to show imprints of Alfvénic waves in the photosphere and chromosphere, respectively. Together with the rotational speeds reported for chromospheric swirls, our study strongly suggests a strong connection between coherent vortical structures in the solar atmosphere and Alfvénic waves. 

In conclusion, this paper demonstrates the reliability and the capability of the SWIRL algorithm in identifying vortical motions in magnetized, turbulent, and highly dynamical astrophysical flows such as those characterizing the solar atmosphere. We believe that, combined with state-of-the-art methods for the estimation of horizontal velocities and high-resolution observational campaigns, the SWIRL algorithm can provide reliable information from which  rigorous conclusions can be drawn as to the statistical properties and nature of swirls in the solar atmosphere.  

%
%

\begin{acknowledgements}
The authors acknowledge support by Swiss National Science Foundation under grant ID 200020\_182094 and the University of Zürich under grant UZH Candoc 2022 ID 7104. This work has profited from discussions with the team of K.\,Tziotiou and E.\,Scullion (conveners) ``The Nature and Physics of Vortex Flows in Solar Plasma'' and with the team of P. Keys (convener) ``WaLSA: Waves in the Lower Solar Atmosphere at High Resolution'' (\url{www.walsa.com}), 
both at the International Space Science Institute (ISSI). 
\end{acknowledgements}


%
%

\bibliographystyle{aa} 
\bibliography{biblio.bib} 

%
%

\begin{appendix}



\section{Modeling chromospheric swirls}
\label{app:A}

\begin{figure}
        \centering
        \resizebox{\hsize}{!}{\includegraphics{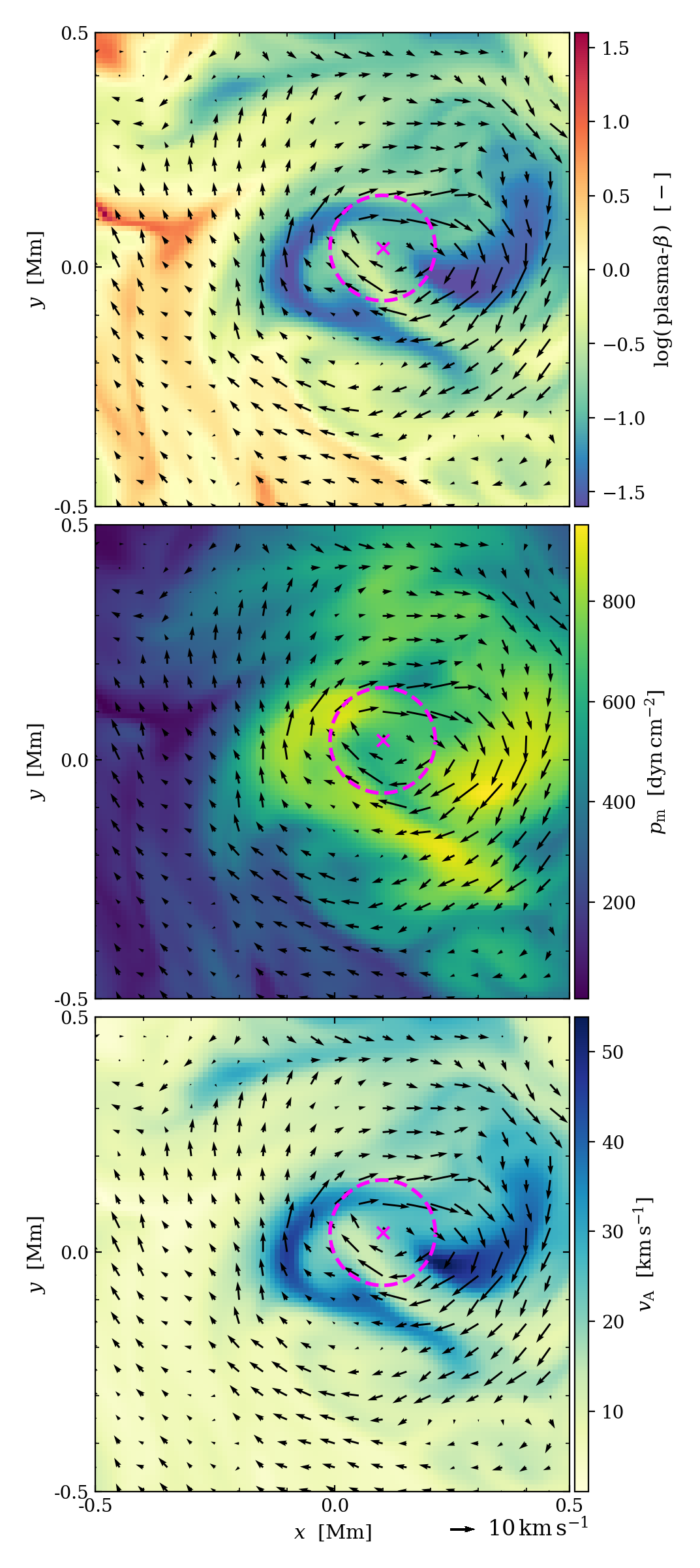}}
        \caption{Physical conditions in the chromospheric horizontal section ($z=700\,{\rm km}$) of the isolated vortical structure visualized in Fig.\,\ref{fig:test_sim_vortexB}. The plasma-$\beta$ (top), magnetic pressure, $p_{\rm m}$ (middle), the local Alfvén speed, $v_{\rm A}$ (bottom) are shown. The horizontal velocity field is represented by a vector field, while the purple circle denotes the position and effective size of the chromospheric swirl.}
        \label{fig:appendix_A}
\end{figure}

We show that the model proposed in Sect.\,\ref{subsubsec:models_of_chromospheric_swirls} for chromospheric swirls is qualitatively consistent with the swirling structures observable in the simulation. We chose to analyze the swirl shown in the three-dimensional rendering of Fig.\,\ref{fig:test_sim_vortexB}, because it represents a nearly ideal case of an isolated swirl. We recall that our model consists of a magnetic flux tube that swirls as a rigid body and is characterized by plasma-$\beta \lesssim 1$. We find that the magnetic flux tube (or vortex) center must correspond to a minimum in magnetic pressure and that the rotational velocity is comparable to the local Alfvén speed, $v_{\rm A}$. 

The top panel of Fig.\,\ref{fig:appendix_A} shows the plasma-$\beta$ in a $1.0\times1.0\,{\rm Mm}^2$ section centered on the swirl at the height of $700\,{\rm km}$, that is, at the bottom of the chromosphere. The swirling plasma is depicted through the vector plot. The size of the chromospheric swirl is probably underestimated by the SWIRL algorithm. We notice that the plasma-$\beta$ is consistently lower than $1$ in the vortical region, with minimal values reaching plasma-$\beta \sim 0.05$ in the outer layers of the swirl. Therefore, the dynamics of the swirl under consideration are mostly dominated by the magnetic field, in agreement with our assumption. 

The magnetic pressure, $p_{\rm m}$, is shown in the middle panel of Fig.\,\ref{fig:appendix_A}. The swirling plasma is highly magnetized, while regions further away show a very low magnetic pressure. Nonetheless, the vortex core is found in a local minimum of magnetic pressure, which means that a negative pressure gradient ---which opposes the centrifugal force--- is directed towards the center of the swirl. Finally, we show in the bottom panel of the same figure that the rotational speed of the chromospheric swirl ($\sim 1\,{\rm km}\,{\rm s}^{-1}$) is of approximately the same magnitude as the local Alfvén speed ($\sim 1-4\,{\rm km}\,{\rm s}^{-1}$). Moreover, we also find a local minimum of the Alfvén speed in proximity to the vortex core, which corroborates Eq.\,(\ref{eq:model_relation}) and our findings regarding Fig.\,\ref{fig:Test_Stat_BrR_relation}.

In conclusion, the simple model of a chromospheric swirl embedded in a strong magnetic flux tube we put forward in Sect.\,\ref{subsubsec:swirls_and_magnetic_fiels} seems to qualitatively capture the main physical properties of an isolated vortical structure that self-consistently emerged from the CO5BOLD simulation. 

\end{appendix}
%
%
\end{document}